\documentclass[12pt]{scrartcl}

\usepackage{amsmath}

\usepackage{cite}
\usepackage{pslatex}
\usepackage{graphicx}
\usepackage[latin1]{inputenc}
\usepackage[T1]{fontenc}

\graphicspath{{plots/}}

\usepackage{color}
\usepackage{colordvi}

\definecolor{mygreen}{rgb}{0,0.5,0}

\def\entry#1#2{\parbox[t]{5.5cm}{\it %
    #1:}\hspace*{0.35cm}\parbox[t]{10.0cm}{#2}\\[-0.2cm]}

\def\ngluon{{\tt NGluon}}
\def\eps{\epsilon}

\def\Eq#1{Eq.~(\ref{#1})}
\def\Fig#1{Fig.~\ref{#1}}
\def\Tab#1{Tab.~\ref{#1}}
\def\Ref#1{Ref.~\cite{#1}}
\def\Refs#1{Refs.~\cite{#1}}

\begin{document}

\begin{titlepage}
\noindent
DESY 10-200 \\
HU-EP-10/74\\
SFB/CPP-10-109
\vspace{1.3cm}

\begin{center}
  {\large\bfseries NGluon: A Package to Calculate One-loop Multi-gluon Amplitudes
    }
  \vspace{1.5cm}

  {\large
  S.~Badger$^{\,a}$, B.~Biedermann$^{\,b}$ and P.~Uwer$^{\,b}$}\\
  \vspace{1.2cm}

  {\it 
    $^a$Deutsches Elektronensynchrotron DESY \\
    Platanenallee 6, D--15735 Zeuthen, Germany\\
    \&\\
    The Niels Bohr International Academy and Discovery Center \\
    The Niels Bohr Institute\\
    Blegdamsvej 17, DK-2100 Copenhagen, Denmark

    \vspace{0.2cm}

    $^b$Humboldt-Universit\"at zu Berlin, Institut f\"ur Physik\\
    Newtonstra{\ss}e 15, D-12489 Berlin, Germany
    }
  \vspace{1.4cm}

  {\large\bf Abstract}
  \vspace{0.5cm}

  \parbox{12.5cm}{%
    We present a computer library for the numerical evaluation of colour-ordered $n$-gluon
    amplitudes at one-loop order in pure Yang-Mills theory. The library uses the recently developed
    technique of {\it generalised unitarity}. Running in double precision the library yields
    reliable results for up to 14 gluons with only a small fraction of events requiring a
    re-evaluation using extended floating point arithmetic. We believe that the library presented
    here provides an important contribution to future LHC phenomenology. The program may also prove
    useful in cross checking results obtained by other methods. In addition, the code provides a
    sample implementation which may serve as a starting point for further developments.
    }    
\end{center}
\end{titlepage}

%
%

\section*{Program summary}
\entry{Title of program}{{\tt NGluon}}

\entry{\it Version}{1.1}

\entry{Catalogue number}{}

\entry{Program summary URL}{\tt
  http://www.physik.hu-berlin.de/pep/tools \\
  }

\entry{E-mail}{\tt Simon.Badger@nbi.dk,\\
  Benedikt.Biedermann@Physik.HU-Berlin.de,\\
  Peter.Uwer@Physik.HU-Berlin.de
  }

\entry{License}{GNU Public License}

\entry{Computers}{Any computer platform supported by the GNU compiler suite.}

\entry{Operating system}{---no specific requirements---, tested on
  Scientific Linux 5.2}

\entry{Program language}{C++}

\entry{Memory required to execute}{Depending on the complexity, for
  realistic applications like 10 gluon production in double precision
  below 10 MB}

\entry{Other programs called}{---none---}

\entry{External files needed}{ QCDLoop, qd}

\entry{Keywords}{unitarity method, one-loop corrections}

\entry{Nature of the physical problem}{Evaluation of next-to-leading
  order corrections for gluon scattering amplitudes in pure gauge theory.}

\entry{Method of solution}{Purely numerical approach based on tree
  amplitudes obtained via Berends-Giele recursion combined with
  unitarity method}

\entry{Restrictions on complexity of the problem}{Running in double
  precision the number of gluons should not exceed 14}

\entry{Typical running time}{Depending on the number of external
  gluons between less than a milli second (4 gluons) up to a 1s (14
  gluons) per phase space point.}

\newpage
%
%
\section{Introduction}
\label{sec:intro}
The Large Hadron Collider at CERN allows the exploration of a complete new
energy regime and will help us to unravel the mechanism of
electroweak symmetry breaking. However, the large QCD background to essentially
all major signal processes makes any potential discovery at
the LHC a highly non-trivial endeavour. A necessary prerequisite is
thus a solid understanding of the QCD backgrounds. This includes
sophisticated methods to determine the background from data but also
improved theoretical calculations providing reliable
predictions. Leading-order predictions in QCD are usually plagued by large
uncertainties due to the residual scale dependence. For reliable
predictions higher order corrections, in particular next-to-leading
order (NLO) calculations, are mandatory. With an increasing number of
particles involved in the hard scattering process the evaluation of
the corresponding one-loop amplitudes becomes more and more complicated. 
In recent years considerable progress has been made towards a fully automated procedure for
NLO corrections to perturbative QCD cross sections. The virtual corrections
to multi-particle amplitudes were for a long time considered to be the bottleneck in multi-jet cross
sections predictions for high energy hadron collisions at the Tevatron and LHC.
Over the past 15 years essentially two methods have been used. One is
the traditional approach based on the evaluation of Feynman
diagrams. In this approach the large number of Feynman diagrams and
related to that the increasing algebraic complexity may actually put
limitations on the processes which are feasible following this
technique. Also numerical stability and speed are non-trivial
issues. However despite these problems many important results have
been obtained along these lines  (see for example
\cite{Bredenstein:2009aj,Dittmaier:2007th,Dittmaier:2007wz},
we refer to \Ref{Binoth:2010ra} for a more complete review of the current status). The second
method makes use of unitarity and in its original version tries to
reconstruct the loop amplitudes via the Cutkosky rules. The first
applications of this method to jet physics date back to the mid
nineties \cite{Bern:1994zx,Bern:1994cg}. At that time the method
was used only by very few groups. This situation has changed dramatically in the
past five years and   
the method of using unitarity cuts to construct one-loop gauge theory amplitudes is by now well
established. In more recent years, thanks to a deeper understanding
in the role of complex analysis, the procedure has been generalised to incorporate multiple cuts
\cite{Britto:2004nc} effectively reducing the computation of one-loop
amplitudes to an algebraic procedure where the only input from the underlying field
theoretical model is provided by the respective Born amplitudes. 
For a detailed description of various aspects of this approach we
refer to the vast literature on the subject 
\cite{Ossola:2006us,Forde:2007mi,Bern:2005hs,Bern:2005ji,Bern:2005cq,Berger:2006ci,Berger:2006vq,Anastasiou:2006jv,Anastasiou:2006gt,Giele:2008ve,Ossola:2008xq,Badger:2008cm}.

In this work we follow the algorithm of $D$-dimensional generalised unitarity
\cite{Ellis:2007br,Giele:2008ve} which is closely
related to the integrand reduction of Ossola, Papadopoulos and Pittau (OPP) \cite{Ossola:2006us}. 
This procedure has been implemented successfully in a number of independent, private, codes
\cite{Berger:2008sj,Giele:2008bc,Giele:2009ui,Lazopoulos:2008ex,Lazopoulos:2009zn} that have been
recently applied to phenomenological NLO QCD studies (see for example Refs.
\cite{Berger:2009ep,Berger:2009zg,Berger:2010vm,Berger:2010zx,Melia:2010bm,%
Melnikov:2010iu,Melnikov:2009wh,Melnikov:2009dn,Ellis:2009zw,Bevilacqua:2010ve,Bevilacqua:2009zn}).
In addition two public codes implementing the OPP integrand reduction procedure have been released
\cite{Ossola:2007ax,Mastrolia:2010nb}.

This article is organised as follows. In section \ref{sec:methods} we give a brief overview of the
on-shell techniques implemented in the {\tt NGluon} {\tt C++} package. In section \ref{sec:installation} we describe how to
install the package from the source files.
A short description  of the various public member functions is
presented in section \ref{sec:description}. 
In section \ref{sec:usage} we give some basic examples on how to use the
package and show a detailed analysis on the performance in terms of 
speed and accuracy before
reaching the conclusions in section \ref{sec:conclusions}. 

\section{Methods}
\label{sec:methods}
Since the method has been described in detail in the literature, in this section we present a basic
overview of the generalised unitarity procedure focusing on the algorithm employed in \ngluon.
We restrict the discussion to the purely massless case throughout.
Owing to the choice of the Van Neerven-Vermaseren basis for the loop momenta, our
implementation most closely resembles that used by
the {\tt Rocket} collaboration \cite{Ellis:2007br,Giele:2008ve}.

We split the one-loop gluon amplitudes $A_n^{(1)}$ into two
contributions:
\begin{equation}
  A_n^{(1)} = A_n^{(1),cc} + R_n^{(1)},
  \label{eq:A1}  
\end{equation}
The cut-constructible part $A_n^{(1),cc}$, which
contains all logarithms and divergences, may be computed using four-dimensional cuts. The remaining
rational terms $R_n^{(1)}$ must be extracted using additional information from cuts in $4-2\eps$ dimensions.
It is well known that the cut-constructible part can be written in
terms of a basis of scalar integrals with a maximum of four propagators. Restricting ourselves to the
case of massless propagators we write the cut-constructible term as, 
\begin{equation}
  A_n^{(1),cc} = 
  \sum_{i,j,k,l} C_{4;i|j|k|l}\, I_{4;i|j|k|l}
  + \sum_{i,j,k} C_{3;i|j|k}\, I_{3;i|j|k}
  + \sum_{i,j} C_{2;i|j}\, I_{2;i|j},
  \label{eq:A1cc}
\end{equation}
where $I_4, I_3$ and $I_2$ denote the scalar four-, three- and
two-point one-loop integrals. 
Denoting the set of external momenta as $\{p_i\}$, $i=1,n$, we label the possible internal propagators as:
\begin{equation}
  P_i = \frac{1}{ D_i} = \frac{1}{ (\ell-q_i)^2},
  \label{eq:prop}
\end{equation}
with an integer $i$, where $q_i=\sum_{m=0}^{i} p_m$. In this notation we take $p_0=0$.
The scalar integrals are then given by the collection of propagators as specified by the second multi-index, i.e.
\begin{equation}
  I_{4;i|j|k|l} = \int \frac{d^D\ell}{(2\pi)^D}\, \frac{1}{D_iD_jD_kD_l}.
\end{equation}
For QCD processes these integrals are known in the framework of dimensional regularisation
\cite{'tHooft:1978xw,Denner:1991qq,Bern:1993kr,Ellis:2007qk} and have been more recently made available in a number of public codes {\tt
FF/QCDloop} \cite{vanOldenborgh:1989wn,Ellis:2008qc}, {\tt LoopTools} \cite{Hahn:1998yk} and {\tt
OneLOop} \cite{vanHameren:2010cp}. Therefore, the only process dependent information in \Eq{eq:A1cc} are the
rational coefficients, $C_4,C_3$ and $C_2$.

The rational part $R_n$ can be derived by taking the $\eps\to 0$ limit
of the expanded integral basis in higher dimensions \cite{Giele:2008ve}. After terms of higher order
in $\eps$ are discarded we can write,
\begin{align}
  R_n^{(1)} &= -\frac{1}{6}\sum_{i,j,k,l} C^{[4]}_{4;i|j|k|l}
  - \frac{1}{2} \sum_{i,j,k} C^{[2]}_{3;i|j|k}
  - \sum_{i,j} \frac{s_{i,j-1}}{6} C^{[2]}_{2;i|j}.
  \label{eq:A1rat}
\end{align}
As we will discuss later the super-scripts correspond to the polynomial structure of the $D$-dimensional
integrands. We choose to extract the values of these coefficients by performing the cuts in
four dimensions with an internal mass carrying the $D$-dimensional information. The
coefficients are then computed from the large mass limit of the four-dimensional case. 
This is a numerical translation of the method described in Ref.
\cite{Badger:2008cm} which has also been used in the recent computation of $W/Z+3j$ and
$W+4j$ by the {\tt BlackHat} collaboration \cite{Berger:2009zb,Berger:2009ep,Berger:2010vm,Berger:2010zx}. Alternatively one can extract the same coefficients by
interpolating the result of computations in higher integer dimensions as described by Giele, Kunszt
and Melnikov \cite{Giele:2008ve}.

We follow a top-down approach, starting with the leading singularity coming from box contributions,
working through the triangles to the bubbles. At each stage all possible configurations of propagators
are put on-shell where the one-loop amplitude factorises into products of tree level amplitudes. Knowledge of
the amplitude coming from higher cuts is then systematically removed in such a way that the integral
coefficients can be uniquely identified and their value determined using purely algebraic methods.

\subsection{Universal pole structure}
The poles in the dimensional regularisation parameter, $\eps$, have a universal structure which, for
the case of massless QCD, has been known for some time, see for example \cite{Giele:1991vf}. This
serves as an internal cross check for our computation and takes an extremely simple form for the
colour-ordered gluon amplitudes considered in this paper,
\begin{equation}
  A^{(1)}_n(\{p_k\}) |_{\rm poles} = \left[ \frac{11}{3\eps}
  -\frac{1}{\eps^2} \sum_{i=1}^n \left( \frac{\mu_R^2}{-s_{i,i+1}} \right)^\eps
  \right] A^{(0)}_n(\{p_k\}),
  \label{eq:poles}
\end{equation}
where $s_{i,i+1} = (p_i+p_{i+1} )^2$ and $\mu_R$ is the regularisation scale.

\subsection{Tree-level amplitudes}

The main ingredient for the construction of the one-loop amplitude is an efficient evaluation of the
tree-level amplitudes entering each cut. We have chosen to implement Berends-Giele recursion
relations \cite{Berends:1987me} for the gluonic tree amplitudes and the amplitudes with a pair of massive scalars relevant
for the rational terms. This gives us the benefit of polynomial growth in speed with the number of
gluons. The computational costs for the evaluation of pure gluon amplitudes scales like $n^4$ where $n$ is the number of gluons.
In our implementation we cache all the sub-currents related to the evaluation of the Born amplitude for the process under
consideration. Using this cache the evaluation of every individual sub-amplitude required for the computation of the one-loop amplitudes
scales then like $n^3$. We will come back to this issue when we discuss the overall behaviour.

\subsection{Cut integrals and the loop momentum parametrisation}

\begin{figure}[!ht]
  \begin{center}
    \includegraphics[width=0.7\textwidth]{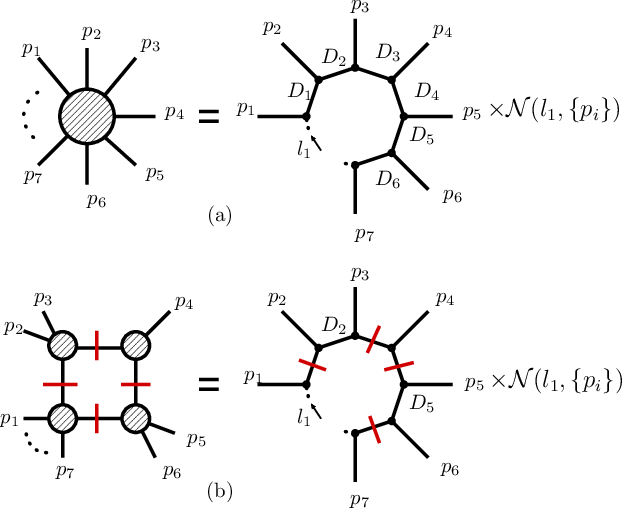}
  \end{center}
  \label{fig:ncut}
  \caption{Figure 1(a) shows the labelling of the propagators in the integral representation of the
  amplitude. Figure 1(b) shows a generic quadruple cut where propagators $D_1,D_3,D_4$
  and $D_6$ are replaced by on-shell delta functions and the amplitude factorises into a product of
  four tree amplitudes.}
\end{figure}

We choose to compute the four-dimensional complex momenta appearing as solutions to the cut
constraints using the van-Neerven-Vermaseren basis \cite{vanNeerven:1983vr} as described in
\Refs{Ellis:2007br,Giele:2008ve}. An alternative approach is to use a Weyl spinor basis as used by
{\tt Blackhat}\cite{Berger:2008sj} and within the OPP approach of Ref.~\cite{Ossola:2007bb}.

A complete discussion of the method has been clearly presented in 
\Refs{Ellis:2007br,Giele:2008ve}. Here we review the essential details
specific to our implementation.
A general representation of the cut integrals appearing in $n$-particle one-loop processes is,
\begin{align}
  A_n^{(1)}(\{p_i\})\big|_{k-cut} = \int \frac{d^D\ell_1}{(4\pi)^D} \left(\prod_{i=1}^k 
                \delta( D_i ) D_i\right)
                \frac{\mathcal{N}(\ell_1,\{p_i\})}{\prod_{j=1}^{n} D_j},
\end{align}
where the propagators and external momenta are defined in eq. (\ref{eq:prop}).
The general loop momentum parametrisation for the $k$-particle cut in four dimensions is then written as
\begin{equation}
 \ell^\mu_{1;k}=V_k^\mu+\sum_{i=1}^{5-k}\alpha_i n_i^\mu,
\end{equation}
where the unit vectors $n_i^\mu$ span the trivial space orthogonal to the physical space vector $V_k^\mu$:
\begin{align}
 V_k\cdot n_j &= 0 , & n_i\cdot n_j &= \delta_{ij},
\end{align}
where $i,j=k\ldots4$. The physical space vector $V_k^\mu$ can be determined from the external momenta $\{p_i\}$.

\subsection{Cut-constructible contributions}

\subsubsection{The box coefficients}

The first step is to compute the leading singularity of the one-loop amplitude coming from the
quadruple cuts. We construct all possible ways to cut four propagators in the amplitude, in each case
using the on-shell delta functions to freeze the loop integration. The general solution to the
on-shell constraints yields two complex solutions and the final result for the box coefficient is
simply the sum over the product of four trees evaluated at each solution \cite{Britto:2004nc}. We first denote the integrand,
summed over internal helicities, as
\begin{align}
  \overline{C}_{4;i|j|k|l} = \sum_{h_1,h_2,h_3,h_4} &
        A^{(0)}(-\ell_1^{-h_1},K_{i,j-1},\ell_2^{h_2})\,
        A^{(0)}(-\ell_2^{-h_2},K_{j,k-1},\ell_3^{h_3})\nonumber\\&\times
        A^{(0)}(-\ell_3^{-h_3},K_{k,l-1},\ell_4^{h_4})\,
        A^{(0)}(-\ell_4^{-h_4},K_{l,i-1},\ell_1^{h_1}),
  \label{eq:boxproduct}
\end{align}
where $K_{m,n-1}$ denote sums over external momenta defined as $K_{m,n-1}=\sum_{i=m}^{n-1} p_i$.
The integrand has a polynomial structure depending on a vector $n_1$ in the trivial space
satisfying
$K_{i,j-1}\cdot n_1 = 0,\,K_{j,k-1}\cdot n_1 = 0,\,K_{k,l-1}\cdot n_1 = 0$ and $n_1\cdot n_1=1$,
\begin{align}
  \overline{C}_{4;i|j|k|l}(\ell_1)  = C_{4;i|j|k|l} + \tilde{C}_{4;i|j|k|l}\, n_1\cdot \ell_1.
\end{align}
It has been shown in Ref. \cite{Ellis:2007br} that
\begin{align}
  C_{4;i|j|k|l} &=
  \frac{1}{2}\left(\overline{C}_{4;i|j|k|l}(\ell_1^+)+\overline{C}_{4;i|j|k|l}(\ell_1^-)\right)\\
  \tilde{C}_{4;i|j|k|l} &=
  \frac{1}{2\sqrt{\Delta_4}}\left(\overline{C}_{4;i|j|k|l}(\ell_1^+)-\overline{C}_{4;i|j|k|l}(\ell_1^-)\right),
\end{align}
where the Gram determinant $\Delta_4$ can be written in terms of the momenta entering the
propagators as,
\begin{align}
  \Delta_4 &= {\rm det}( \bf G) , & G_{ij} &= q_i\cdot q_j.
\end{align}
It is a convenient feature of the van Neerven-Vermaseren basis that $\Delta_4$ and $n_1$ appear
naturally in the construction of the on-shell loop momenta.

\subsubsection{The triangle coefficients}

With only three on-shell constraints, the loop integration of the triple cut has a single degree of
freedom. Building upon the work of \cite{delAguila:2004nf}, it was
shown in \Ref{Ossola:2006us} that by
parameterising the integrand using
unit vectors $n_1$ and $n_2$ spanning the trivial space, the
unknown information can be
extracted as the solution to a system of linear equations. Forde then elegantly demonstrated that one can
use simple complex analysis to show how the box contributions untangle from the triple cut and a
subtracted integrand with polynomial behaviour leads directly to the scalar triangle
coefficient\cite{Forde:2007mi}. Following the construction of \cite{Ellis:2007br}, we implement this procedure
by writing the integrand as,
\begin{align}
  \overline{C}_{3;i|j|k} =& \sum_{h_1,h_2,h_3} 
  A^{(0)}(-\ell_1^{-h_1},K_{i,j-1},\ell_2^{h_2})\,
  A^{(0)}(-\ell_2^{-h_2},K_{j,k-1},\ell_3^{h_3})\,
  A^{(0)}(-\ell_3^{-h_3},K_{k,l-1},\ell_1^{h_1})\nonumber\\&
  - \sum_{l}\frac{\overline{C}_{4;i|j|k|l}(\ell_1)}{(\ell_1-K_l)^2}.
  \label{eq:triproduct}
\end{align}
The polynomial structure in the trivial space can then be written \cite{Ellis:2007br}:
\begin{align}
  \overline{C}_{3;i|j|k}(\ell_1)  &= 
                            C^{(0)}_{3;i|j|k}
                          + C^{(1)}_{3;i|j|k} \alpha_1
                          + C^{(2)}_{3;i|j|k} \alpha_2
                          + C^{(3)}_{3;i|j|k} (\alpha_1^2-\alpha_2^2)\nonumber\\&
                          + C^{(4)}_{3;i|j|k} \alpha_1\alpha_2
                          + C^{(5)}_{3;i|j|k} \alpha_1^2\alpha_2
                          + C^{(6)}_{3;i|j|k} \alpha_1\alpha_2^2,
\end{align}
where $\alpha_k = n_k\cdot \ell_1$, $k=1,2$. The scalar triangle coefficient is
simply $C_{3;i|j|k}=C^{(0)}_{3;i|j|k}$. We extract the coefficients $C^{(m)}_{3;i|j|k}$ using a
discrete Fourier projection.

\subsubsection{Bubble coefficients}

The construction of the scalar bubble coefficients is analogous with the triangle case considered
above. We construct the integrand by subtracting the relevant combination of triangle and box
coefficients from the double cut.
\begin{align}
  \overline{C}_{2;i|j} =& \sum_{h_1,h_2} 
  A^{(0)}(-\ell_1^{-h_1},K_{i,j-1},\ell_2^{h_2})\,
  A^{(0)}(-\ell_2^{-h_2},K_{j,i-1},\ell_1^{h_1})\nonumber\\&
  - \sum_{k}\frac{\overline{C}_{3;i|j|k}(\ell_1)}{(\ell_1-K_k)^2}
  - \frac{1}{2}\sum_{k,l}\frac{\overline{C}_{4;i|j|k|l}(\ell_1)}{(\ell_1-K_k)^2 (\ell_1-K_l)^2}.
  \label{eq:bubproduct}
\end{align}
There are now three vectors spanning the trivial space so the integrand can be written in terms of
nine independent coefficients \cite{Ellis:2007br},
\begin{align}
  \overline{C}_{2;i|j} =& 
         C^{(0)}_{2;i|j}
        +C^{(1)}_{2;i|j}\alpha_1
        +C^{(2)}_{2;i|j}\alpha_2
        +C^{(3)}_{2;i|j}\alpha_3
        +C^{(4)}_{2;i|j}(\alpha_1^2-\alpha_3^2)\nonumber\\&
        +C^{(5)}_{2;i|j}(\alpha_2^2-\alpha_3^2)
        +C^{(6)}_{2;i|j}\alpha_1\alpha_2
        +C^{(7)}_{2;i|j}\alpha_1\alpha_3
        +C^{(8)}_{2;i|j}\alpha_2\alpha_3,
        \label{eq:bubpoly}
\end{align}
where $\alpha_k = n_k\cdot \ell_1$, $k=1,3$. For the massless amplitudes considered here we do not need
to proceed to reduce further and extract the
tadpole coefficients. The computation of $C_{2;i|j} = C^{(0)}_{2;i|j}$ completes the calculation of the
cut-constructible terms. Again we use the discrete Fourier projection to efficiently compute the
coefficients.

\subsection{Rational Contributions}

Using a super-symmetric decomposition of the gluonic loop
\cite{Bern:1994cg} one can show that the rational terms for
our one-loop amplitude are the same as those coming from contributions with a scalar loop. The
information coming from the $4-2\eps$-dimensional cuts can be encapsulated by adding a
mass parameter to the four-dimensional loop momenta,
\begin{equation}
  \ell_{[4-2\eps]} = \ell_{[4]}+\ell_{[-2\eps]},
\end{equation}
where $\ell_{[-2\eps]}^2=-\mu^2$. We then proceed to extract the
coefficients of \Eq{eq:A1rat} 
from cuts with a massive scalar running inside the loop
\cite{Bern:1996ja,Anastasiou:2006gt,Anastasiou:2006jv,Badger:2008cm,Berger:2009zb}. We note that the
supersymmetric decomposition relates the rational part of the gluon amplitude to that of a complex
scalar. Therefore, in order to match with \Eq{eq:A1rat}, the products of trees are all
multiplied by a factor of two. 

\subsubsection{The pentagon coefficients}

An additional complication in the numerical computation of the $D$-dimensional pieces is the
presence of a non-zero pentagon coefficient. Although such a contribution will vanish explicitly in
an analytical calculation here we are forced to include them to ensure a numerically stable result.
There is no trivial space for this contribution and the result appears solely as a subtraction term
for the box coefficient. The coefficient is then simply \cite{Giele:2008ve}, 
\begin{align}
  &R_{5;i|j|k|l} = 
        2A^{(0)}_S(-\ell_1,K_{i,j-1},\ell_2)\,
        A^{(0)}_S(-\ell_2,K_{j,k-1},\ell_3)\nonumber\\&\times
        A^{(0)}_S(-\ell_3,K_{k,l-1},\ell_4)\,
        A^{(0)}_S(-\ell_4,K_{l,m-1},\ell_5)\,
        A^{(0)}_S(-\ell_5,K_{m,i-1},\ell_1).
  \label{eq:pentproductR}
\end{align}
The five on-shell constraints can be satisfied by fixing the four
coefficients of the loop momentum using,
\begin{equation}
  \{
  2\ell_1\cdot K_{i,j-1}=K_{i,j-1}^2,
  2\ell_1\cdot K_{i,k-1}=K_{i,k-1}^2,
  2\ell_1\cdot K_{i,l-1}=K_{i,l-1}^2,
  2\ell_1\cdot K_{i,m-1}=K_{i,m-1}^2
  \},
\end{equation}
and additionally setting the $D$-dimensional mass by $\mu^2=\ell_1^2$. The implementation in {\tt
NGluon} chooses to implement a solution to these on-shell constraints using two-dimensional
Weyl-spinors along the lines of those used in Refs. \cite{Ossola:2006us,Forde:2007mi,Badger:2008cm}
since we found it more efficient and numerically stable than the van-Neerven basis solution used in
Ref. \cite{Giele:2008ve}. This allows us to avoid the computation of any $4\times4$ determinants.

\subsubsection{The box coefficients}

Since we compute the $D$-dimensional coefficient $C_4^{[4]}$ from four-dimensional massive cuts, the
first part of the calculation proceeds exactly as in the cut-constructible case. The quadruple cut
together with the pentagon subtraction is a polynomial in $\mu^2$ from which we can extract the
coefficients using a discrete Fourier projection.
\begin{align}
  \overline{R}_{4;i|j|k|l} = &
        2A^{(0)}_S(-\ell_1,K_{i,j-1},\ell_2)\,
        A^{(0)}_S(-\ell_2,K_{j,k-1},\ell_3)\nonumber\\&\times
        A^{(0)}_S(-\ell_3,K_{k,l-1},\ell_4)\,
        A^{(0)}_S(-\ell_4,K_{l,i-1},\ell_1)\nonumber\\&
        - \sum_{m} \frac{C_{5;i|j|k|l|m}}{(\ell_1-K_m)^2}.
  \label{eq:boxproductR}
\end{align}
The polynomial form of the integrand is,
\begin{align}
  \overline{R}&_{4;i|j|k|l}(\ell_1(\mu^2)) = 
               R^{(0)}_{4;i|j|k|l} + n_1\cdot \ell_1 R^{(1)}_{4;i|j|k|l} \nonumber\\&
        +\mu^2\left( R^{(2)}_{4;i|j|k|l} + n_1\cdot \ell_1 R^{(3)}_{4;i|j|k|l}\right)
        +\mu^4\left( R^{(4)}_{4;i|j|k|l} + n_1\cdot \ell_1 R^{(5)}_{4;i|j|k|l}\right),
\end{align}
where $C^{[4]}_{4;i|j|k|l} = R^{(4)}_{4;i|j|k|l}$. Performing the four-dimensional extraction three
times is then sufficient to extract all the coefficients. We choose the mass integration to lie
on a circle of radius $\mu_0$,
\begin{equation}
  \mu_m^2 = \mu_0^2 \exp\left( \frac{2\pi i m }{3} \right),
\end{equation}
so the explicit solution becomes
\begin{align}
  R^{(2p)}_{4;i|j|k|l} &= \sum_{m=0}^{5} \frac{1}{2\mu_m^{2p}}
  \left(
  \overline{R}_{4;i|j|k|l}(\ell_1^+(\mu_m^2))+
  \overline{R}_{4;i|j|k|l}(\ell_1^-(\mu_m^2)) 
  \right)\\
  R^{(2p+1)}_{4;i|j|k|l} &= \sum_{m=0}^{5} \frac{1}{2\mu_m^{2p}\sqrt{\Delta_4}}
  \left(
  \overline{R}_{4;i|j|k|l}(\ell_1^+(\mu_m^2))-
  \overline{R}_{4;i|j|k|l}(\ell_1^-(\mu_m^2)),
  \right)
\end{align}
with $p=0,1,2$. It is important to choose the value of the radius of integration, $\mu_0$, such that
the quadruple cut and subtraction terms are of the same order of magnitude. There are various ways
to do this. In {\tt NGluon} the radius is scaled with respect to the largest pentagon contribution that
occurs in the subtractions in order to maximise numerical stability. Since the rank of tensor
integrals is constrained to be of maximum four in gauge theory, it will always be the case that
$R^{(5)}_{4;i|j|k|l}=0$. This can be a useful test of the accuracy of the computation of
$\overline{R}_{4;i|j|k|l}$.
  
\subsubsection{The triangle coefficients}

At this stage the method should be quite clear. We proceed to extract the triangle coefficients
from the massive scalar loop, sampling over the mass parameter to find the coefficient contributing
the the rational term. In order to have an integrand with polynomial behaviour we must subtract both
pentagon and box contributions from the product of trees,
\begin{align}
  \overline{R}_{3;i|j|k} =&
        2A^{(0)}_S(-\ell_1,K_{i,j-1},\ell_2)
        A^{(0)}_S(-\ell_2,K_{j,k-1},\ell_3)
        A^{(0)}_S(-\ell_3,K_{k,i-1},\ell_1)\nonumber\\&
        - \sum_{l} \frac{\overline{R}_{4;i|j|k|l}}{(\ell_1-K_l)^2}
        - \frac{1}{2}\sum_{l,m} \frac{R_{5;i|j|k|l|m}}{(\ell_1-K_l)^2(\ell_1-K_m)^2}.
  \label{eq:triproductR}
\end{align}
The polynomial structure of the integrand can be written:
\begin{align}
  \overline{R}_{3;i|j|k}(\ell_1)  &= 
                            R^{(0)}_{3;i|j|k}
                          + R^{(1)}_{3;i|j|k} \alpha_1
                          + R^{(2)}_{3;i|j|k} \alpha_2
                          + R^{(3)}_{3;i|j|k} (\alpha_1^2-\alpha_2^2)\nonumber\\&
                          + R^{(4)}_{3;i|j|k} \alpha_1\alpha_2
                          + R^{(5)}_{3;i|j|k} \alpha_1^2\alpha_2
                          + R^{(6)}_{3;i|j|k} \alpha_1\alpha_2^2\nonumber\\&
                          + \mu^2\left(
                          R^{(7)}_{3;i|j|k} \alpha_1
                          + R^{(8)}_{3;i|j|k} \alpha_2
                          + R^{(9)}_{3;i|j|k} \right),
\end{align}
where $\alpha_1,\alpha_2$ are identical to that of the cut-constructible triangle. The Fourier
projection over the mass proceeds as in the box contributions except we are required to sample over
more points due to the larger number of independent coefficients.

\subsubsection{The bubble coefficients}

No new features appear in the extraction of this final term in the amplitude so we simply write down
the formulae for the integrand and it's polynomial structure. We point readers towards 
\Refs{Ellis:2007br,Giele:2008ve,Berger:2008sj,Ossola:2008xq,Ossola:2006us} for further details,
\begin{align}
  \overline{R}_{2;i|j} = &
        2A^{(0)}_S(-\ell_1,K_{i,j-1},\ell_2)
        A^{(0)}_S(-\ell_2,K_{j,i-1},\ell_1)   
        - \sum_{k} \frac{\overline{R}_{3;i|j|k}}{(\ell_1-K_k)^2}\nonumber\\&
        - \frac{1}{2}\sum_{k,l} \frac{\overline{R}_{4;i|j|k|l}}{(\ell_1-K_k)^2(\ell_1-K_l)^2}
        - \frac{1}{6}\sum_{k,l,m}
        \frac{R_{5;i|j|k|l|m}}{(\ell_1-K_k)^2(\ell_1-K_l)^2(\ell_1-K_m)^2},
  \label{eq:bubproductR}
\end{align}
and
\begin{align}
  \overline{R}_{2;i|j} =& 
         R^{(0)}_{2;i|j}
        +R^{(1)}_{2;i|j}\alpha_1
        +R^{(2)}_{2;i|j}\alpha_2
        +R^{(3)}_{2;i|j}\alpha_3
        +R^{(4)}_{2;i|j}(\alpha_1^2-\alpha_3^2)\nonumber\\&
        +R^{(5)}_{2;i|j}(\alpha_2^2-\alpha_3^2)
        +R^{(6)}_{2;i|j}\alpha_1\alpha_2
        +R^{(7)}_{2;i|j}\alpha_1\alpha_3
        +R^{(8)}_{2;i|j}\alpha_2\alpha_3
        + \mu^2 R^{(9)}_{2;i|j}.
\end{align}
$\alpha_i$ are the same as those in \Eq{eq:bubpoly} and the coefficient of eq.
\Eq{eq:A1rat} is $C^{[2]}_{2;i|j} = R^{(9)}_{2;i|j}$.

%
%
\section{Installation}
\label{sec:installation}
\def\dirname{{\tt NGluon-1.1}~} 
\def\objs#1{{\tt /obj#1}} 
{\tt NGluon} uses the GNU compiler suite and is available as a tarball, {\tt
NGluon-1.1.tar.gz}, from\\
\verb#http://www.physik.hu-berlin.de/pep/tools#.\\
If {\tt NGluon} is used without {\tt QCDLoop} and without the {\tt qd} extension the g++ compiler 
is sufficient to compile and install the package. To do this first unpack the archive using:
\begin{verbatim}
tar xvfz NGluon-1.1.tar.gz
\end{verbatim}
You can then move to the directory \dirname and type {\tt make}. This will build the
{\tt NGluon} library (for static linking): {\tt libNGluon.a}. 
The object files and library are placed into
the directory \dirname\objs{}. In addition an example application {\tt
NGluon-demo} will be created. We note that most of the files found in \dirname belong to the sample
application. In particular these files contain additional code to
generate phase space points and analytic results.
For the details we refer to \Tab{tab:filelist} where a
short description of the files is given. The upper block of the table describes
the files necessary to build the library. Files in the lower block are needed
only for the example applications. We will not describe files in the lower block
in detail since they are only provided for illustrative purpose
and are not part of the {\tt NGluon} package itself.

\begin{table}[htbp]
  \begin{center}
    \leavevmode
    \begin{tabular}[h]{l|l}
      \multicolumn{2}{c}{NGluon library}\\
      \hline
      File name & Functionality\\
      \hline
      NGluon.h  & Header file for NGluon\\
      NGluon.cpp  & Source file with the implementation of the
      unitarity method \\
      Current.h  & Header file for Berends-Giele related functions \\
      Current.cpp  & Implementation of the Berends-Giele recursion \\
      Coefficients.h  & Definition of storage used internally\\
      LoopIntegrals.cpp  & Interface to {\tt QCDLoop}\\
      LoopIntegrals.h  & Interface to {\tt QCDLoop}\\     
      mytypes.h  & Header file to switch to extended precision using
      {\tt qd}
      \\
      \hline 
      \multicolumn{2}{c}{}\\
      \multicolumn{2}{c}{Sample application}\\
      \hline
      File name & Functionality\\
      \hline
      analytic.h & Header file for analytic formulae\\
      analytic.cpp & Implementation of some analytic formulae \\
      NGluon-demo.cpp & Example application\\
      GKM.cpp & Results from \Ref{Giele:2008ve}\\
      GKM.h & Results from \Ref{Giele:2008ve}\\
      GZ.cpp & Results from \Ref{Giele:2008bc}\\
      GZ.h & Results from \Ref{Giele:2008bc}\\
      FourMomentum.h & Four momentum class\\
      histogram.h & Simple histogram functionality\\
      phasespace.cpp & Phase space generation\\
      phasespace.h & Phase space generation\\
      \hline 
      \multicolumn{2}{c}{}\\
      \multicolumn{2}{c}{Make files}\\
      \hline
      File name & Functionality\\
      \hline
      Makefile & Makefile to built the libraries and compile the
      example applications.\\
      Makefile.all & Makefile specific instances of the library 
      and create driectory structure\\
      Makefile.inc & Configuration file for the Makefile\\
    \end{tabular}
    \caption{Files included in the {\tt NGluon} package. In the upper
      part files belonging to the {\tt NGluon} library itself are listed.
      In the lower part files which are only used by the example 
      application {\tt NGluon-demo} are shown.}
    \label{tab:filelist}
  \end{center}
\end{table}

Without a library for the scalar one-loop integrals {\tt NGluon}
calculates the coefficients of the scalar integrals as well as the
rational part. However since all the scalar integrals are set to one
the full result for the full amplitude is meaningless. As a
consequence most of the tests which can be found in {\tt  NGluon-demo.cpp}
will not work. 
To obtain the full functionality {\tt NGluon} should be combined with
a library for the evaluation of the scalar one-loop integrals.
{\tt NGluon} is prepared for use with {\tt FF} \cite{vanOldenborgh:1989wn} and
{\tt QCDLoop} \cite{Ellis:2007qk} for the evaluation of the scalar
integrals. The package has been tested with {\tt QCDLoop-1.9} which can be downloaded at the
following address:\\
\verb#http://qcdloop.fnal.gov/#\\
Note that per default {\tt QCDLoop} uses g77. Since g77 is no longer
supported we recommend to switch to gfortran. If g77 shall be used the
user needs to adopt the makefile in \dirname. In particular the linker
options have to be adjusted to enable the linking of code compiled
with the fortran compiler together with the main program compiled with
g++. 
To use {\tt QCDLoop} the
user should first install the {\tt QCDLoop} library. For the details
how to do this we refer to the {\tt QCDLoop} documentation.



To link {\tt QCDLoop} with {\tt NGluon} it is sufficient to edit the 
configuration of {\tt NGluon} which is controlled via the file: 
{\tt Makefile.inc}. To include {\tt QCDLoop} the variable  {\tt
  ENABLE\_QL} is changed to {\tt ENABLE\_QL=yes}.
In addition the path to the {\tt QCDLoop-1.9} installation (the {\tt
  QCDLoop-1.9} directory) needs to be be configured through the
variable {\tt QLDIR} in {\tt Makefile.inc}. Typing,
\begin{verbatim}
make clean
make
\end{verbatim}
will then compile a version of {\tt NGluon} including the scalar
one-loop integrals from the {\tt QCDLoop} library. 

{\tt NGluon} is also prepared to work with extended floating point
arithmetic as provided for example by the {qd} library \cite{QD}.
The library can be obtained at the following address:\\
\verb#http://crd.lbl.gov/~dhbailey/mpdist/#\\
{\tt NGluon} has been tested with version {\tt qd-2.3.11}. 

Once the {\tt qd} library has been installed one may easily compile quadruple
and octuple precision versions of {\tt NGluon} via {\tt Makefile.inc}. Simply
change {\tt ENABLE\_DD=yes} (``double-double'') or {\tt ENABLE\_QD=yes} 
(``quad-double'')
in the configuration file {\tt Makefile.inc}. The
path to the library and the location of the header files needs to be
configured via {\tt QDLIB} and {\tt QDINCLUDE}. Again,
\begin{verbatim}
make clean
make
\end{verbatim}
will then compile the relevant versions of the {\tt NGluon} library placing the
library and object files into the directories \dirname\objs{}, 
\dirname\objs{-dd}
and \dirname\objs{-qd}. 
Up to three versions of the test program are also created
in \dirname : {\tt NGluon-demo}, {\tt NGluon-demo-dd} and {\tt NGluon-demo-qd}.
Since the object files are put into different directories, the
different versions do not interfere with each other and can be used in 
parallel.
Note that the floating point arithmetic used in {\tt QCDLoop} is not changed.
In particular, if numerical instabilities arise in the evaluation of 
the scalar integrals they would not be cured by switching to extended
precision.

If specific compiler options are required these options must be added also in
{\tt Makefile.inc} via the {\tt CFLAGS} variable. Non-standard locations for
other libraries may be added to {\tt LIBS}, {\tt LFLAGS} and {\tt IFLAGS}.

One can switch the compiler via the standard makefile variable {\tt CXX}.
The degree of optimisation can be changed via {\tt OPT} though we recommend
{\tt -O2}.

%
%
\section{Description}
\label{sec:description}
We decided to encapsulate the entire implementation in a class called
{\tt NGluon}. The main purpose of this approach is to hide most of
the internal data required to store partial results from the user. 
To instantiate an object of the {\tt NGluon} class the following
constructor is used (the only one available):
\begin{verbatim}
NGluon loop_amp(ngluon,moms,helicities);
\end{verbatim}
where {\tt ngluon} is an integer denoting the number of gluons, and {\tt moms} 
specifies a pointer to an array containing the momentum configuration
with the momenta counted outgoing. The corresponding C++ definition would be: 
\begin{verbatim}
DOUBLE moms[ngluon][4].
\end{verbatim}
Note that we use everywhere the preprocessor macro {\tt DOUBLE}
instead of the built-in data type {\tt double}. Using the {\tt qd}
library \cite{QD} this
allows us to create a version of the program using extended floating
point arithmetic by simply recompiling the program.
The header file {\tt mytypes.h} takes care to set the macro {\tt
  DOUBLE } to the required value that is either~\verb#dd_real# or~
\verb#qd_real# when compiling with extended precision or {\tt double }
when built-in double precision shall be used. Instead of the
preprocessor variable to control the data type, we could have used C++ 
templates. However, code generation via the compiler is much harder to
control in this case. In addition this approach often leads to longer
executables which may affect the performance in a negative way.
The last argument in the constructor specifies an integer array where the
helicities ($\pm 1$) for the gluons are stored. 
The corresponding C/C++ definition reads: 
\begin{verbatim}
int helicities[ngluon].
\end{verbatim}
Note that these arrays are not copied by the {\tt NGluon}
class, only the address of the arrays is stored in the {\tt NGluon}
object. After updating the momentum configuration or the helicity
configuration {\tt NGluon} will thus automatically 
use the updated quantities in the next call. We note also that it is not
foreseen to change the number of gluons after the {\tt NGluon} object has
been constructed. To study amplitudes with differing numbers of gluons a new instance must be
constructed for each case. Since all local data is
stored inside the class, these instances do not interfere with each
other. In principle the class itself should also be thread-safe.
 
Below we give a list of all public methods together with
a short description.

\verb#static void setVerbosity(VERBOSITY output_);#\\
This function controls the verbosity of the {\tt NGluon} class. Using 
{\tt NGluon::QUIET} as argument turns all debugging information off while 
{\tt NGluon::FULL} switches to maximal verbosity.

\verb#void setMuR(DOUBLE muR_);# \\
Used to set the renormalisation scale used in the scalar one-loop integrals.
Per default the renormalisation scale is set to 1.

\verb#void setScaleTest(bool scaleTest_);#\\
If the argument is {\tt true} the function switches the scale test
on. For a detailed description see below. We note that if 
the scale test is switched on the runtime doubles, 
however a reliable estimate for the accuracy is 
provided for the final result. By default the scale test is switched
off.

\verb#std::complex<DOUBLE> evalAmp();# \\
Calling this function will evaluate the n-gluon amplitude for the
momentum and helicity configuration provided in the constructor. The
return value is the finite part of the amplitude.

\texttt{%
std::complex<DOUBLE> getAfinite(),\\
std::complex<DOUBLE> getAtree(),\\
std::complex<DOUBLE> getAeps2(),\\
std::complex<DOUBLE> getAeps1(),\\
std::complex<DOUBLE> getAcc(),\\
std::complex<DOUBLE> getArat():\\} 
These functions give access to the finite part of the one-loop
amplitude as well as to individual contributions like the value
of
the corresponding tree amplitude, the $1/\epsilon^2$- and 
$1/\epsilon$-poles as well as the cut-constructible (cc) and the 
rational part (rat).

\texttt{%
std::complex<DOUBLE> getAbsError(),\\
std::complex<DOUBLE> getAbsErrorEps1(),\\
std::complex<DOUBLE> getAbsErrorEps2(),\\
std::complex<DOUBLE> getAbsErrorRR(): }\\
these functions provide an estimate for the absolute uncertainty of individual
contributions. For the details how these estimates are obtained
see the discussion at the end of this section.

\texttt{%
DOUBLE getRelError(),\\
DOUBLE getRelErrorCC(),\\
DOUBLE getRelErrorRR(),\\
DOUBLE getRelErrorEps1(),\\
DOUBLE getRelErrorEps2(): }\\
Similar to the functions described above. However instead of an
estimate for the absolute uncertainty the relative
uncertainty is returned.

\verb#DOUBLE IRpoles(const int eps);#\\
This function returns the IR poles obtained from the analytic formulae
(see \Eq{eq:poles})
for the given momentum and helicity configuration. The return value 
is the pre-factor multiplying the corresponding born amplitude without the
pole itself. For {\tt eps=-2} the $1/\epsilon^2$-pole is returned. For 
{\tt eps=-1} the $1/\epsilon$-pole is returned. 
The result is used as a
cross check of the results obtained by {\tt NGluon} from the direct 
numerical evaluation.

\verb#static void sethel(int ngluon, int htype, int helicity[] );#\\
The function creates specific helicity configurations for {\tt
  ngluon} gluons. The configuration is stored in the array specified
as third argument. More specifically the configurations are:
\begin{center}
\begin{tabular}[h]{|l|l|}
  \hline
  {\tt htype} & configuration \\ \hline
  0 & $(+)^n $ \\ \hline
  1 & $-(+)^{n-1}$ \\ \hline
  2 & $--(+)^{n-2}$\\ \hline
  3 & $(-+)^{n/2}$\\ \hline
  4 & $(+-)^{n/2}$\\ \hline
\end{tabular}  
\end{center}

\verb#static void sethel(int ngluon, std::string hstr, int helicity[]);#\\
The function sets the helicity configuration for {\tt ngluon} gluons
specified through a string in the form {\tt "+-++-..."}.

\verb#static std::string helicity2string(const int ngluon, int hel[]);#\\
The function converts a helicity configuration for {\tt ngluon} gluons
specified through the integer array {\tt  int hel[]} into a
string.

A crucial point in the numerical evaluation of one-loop amplitudes is
the control of numerical uncertainties and instabilities. To assure
the correctness of the calculated scattering amplitudes we need checks
to test the reliability of the results. Where analytic results are
known, we may compare with them and we discuss such
comparisons in the next section. However in most cases analytic
results are not available.
In such cases important information can be
obtained from testing general properties of the amplitudes. Such a
test is for example provided by the evaluation of the IR poles of the
amplitude. The IR poles are analytically known due to the universal
structure of IR phenomena in QCD. We can thus compare
what we obtain from the numerical evaluation with what is predicted by
QCD. This check tests the coefficients of the IR divergent triangle
and box integrals. More precisely a specific linear combination of these
coefficients is checked. 
Similar the UV structure which is also known analytically can be
used as a test. This check provides information on the coefficients of
the two-point integrals.
Testing the remaining parts of the cut-constructible contribution to
the full amplitude as well as the test of the rational part is more 
involved. Useful information may be obtained from the evaluation of the
Fourier projection. It is possible to calculate further terms in the
Fourier projection which are predicted to be zero. One can then check
to what extend the numerical results are compatible with
zero. This gives important information on the accuracy of
the Fourier projection. An important property of this check is that
the additional computing effort is moderate. However, the
interpretation of the result of this check in terms of errors on
specific coefficients might be non-trivial in particular when large 
cancellations between individual coefficients take place. 
Since a solid error estimate is crucial, we developed a simple but
very effective method to get a reliable estimate. It is based on the
simple observation that we can rescale the momenta and recalculate the 
amplitude. Owing to the rescaling we call this test {\it scale
  test}. From a physical point of view the test corresponds to use two 
different units in specifying the momenta, i.e. MeV and GeV for example.
Since we know how the amplitudes scale
when we rescale the momenta it is possible to compare the two results
with each other. Naively, one could expect to get precisely
the same result. This could be true even in the presence of rounding
errors if the scaling would only affect
the exponent of the floating point representation. However, rescaling
with a factor which cannot be absorbed into a shift of the exponent in
the binary representation of the floating point number will lead to a 
different mantissa. The floating point arithmetic thus becomes
different. That is digits in the final result which are
affected by rounding errors or numerical instabilities will change.
That gives us a very simple method to check the reliability
of individual contributions to the amplitude even for contributions where no
analytic results are available. We should mention that this
luxury comes at the price of a doubled runtime since every phase space
point is calculated twice. For dedicated comparisons
between different codes or with analytic formula we feel however that
the effort is well spent. In practical applications one would use the 
scale test only for phase space points where we have indications
--- based on the checks described before --- that the result might be 
unreliable. If the scale test leads to the result that the point is
reliable this procedure is less computing extensive than switching to
extended accuracy which would be done only if the scale test leads to
the conclusion that the accuracy does not meet the requirements.
As described above the test can be switched on and off using the
function {\tt setScaleTest}. We will illustrate the scale test in the
next section. 

%
\section{Usage and examples}
\label{sec:usage}
Before discussing the performance of the implementation, let us first
present a simple example to illustrate how the package is
used. All results of this section were produced using the program 
{\tt NGluon-demo} which is included in the package. The user can thus
easily reproduce the numerical results presented in this article. 
The file {\tt NGluon-demo.cpp} may also serve
in providing further examples how to use {\tt NGluon}.
The following example which we will discuss in detail is taken from 
the routine {\tt GZcheck} in 
{\tt NGluon-demo.cpp}. The routine compares results obtained with {\tt NGluon}
with results given in \Ref{Giele:2008bc}:
{\small\begin{verbatim}
void GZcheck(){

  cout << "\n\n";
  cout << "---------------------------------------------------------------\n";
  cout << "Numerical comparison with values published in: \n";
  cout << "Giele,Zanderighi: \n";
  cout << "On the Numerical Evaluation of One-Loop Amplitudes:\n";
  cout << "The Gluonic Case.\n";
  cout << "JHEP 0806:038,2008. \n";
  cout << "---------------------------------------------------------------\n";
  cout << "\n\n";

  cout.setf(ios_base::scientific, ios_base::floatfield);
  cout.precision(15);
  
  const int nlist[7]={6,7,8,9,10,15,20};
  
  DOUBLE k[20][4];
  int helicities[20];
  int ngluon; 
  
  DOUBLE GZres[5][4];

  for(int ngidx=0; ngidx<7; ngidx++){
    ngluon = nlist[ngidx];
    cout << "-------------------------------"
         <<"--------------------------------\n";
    cout << "#Number of gluons = " << ngluon << endl;
    cout << "-------------------------------"
         <<"--------------------------------\n";
    GZsetmom(ngluon,k,GZres);
    NGluon loop_amp(ngluon,k,helicities);
    loop_amp.setMuR(double(ngluon*ngluon));

    DOUBLE res;
    for(int hidx=0;hidx<5;hidx++){
      NGluon::sethel(ngluon,hidx,helicities);
      cout << "Helicities: " << NGluon::helicity2string(ngluon,helicities) 
           << endl;
      loop_amp.evalAmp();
      cmp("tree     ","(GZ)  ",loop_amp.getAtree(),GZres[hidx][0],res);
      cmp("|Aeps2|  ","(GZ)  ",loop_amp.getAeps2(),GZres[hidx][1],res);
      cmp("|Aeps1|  ","(GZ)  ",loop_amp.getAeps1(),GZres[hidx][2],res);
      cmp("|Afinite|","(GZ)  ",loop_amp.getAfinite(),GZres[hidx][3],res);
    }
  } 
}
\end{verbatim}}
The arrays {\tt DOUBLE k[20][4]} and {\tt int helicities[20]} are used
to store the momentum and helicity configuration as described before. 
We loop over the examples of 6,7,8,9,10,15, and 20 gluons as presented in \Ref{Giele:2008bc}. Inside
the loop we load first the momentum
configuration as specified in \Ref{Giele:2008bc} by calling 
{\tt GZsetmom(ngluon,k,GZres)}. The function {\tt
  GZsetmom(ngluon,k,GZres)} loads also the results as
given in \Ref{Giele:2008bc} (Tab. 1 -- Tab. 7 in \Ref{Giele:2008bc}). The next step is then to create an object of the
{\tt NGluon} class. The renormalisation scale is set to the value
used in \Ref{Giele:2008bc} and the scale test is switched off. It follows a loop
over the different helicity configurations used in \Ref{Giele:2008bc}.
The configurations are set using 
{\tt NGluon::sethel(ngluon,hidx,helicities)}.
By calling {\tt loop\_amp.evalAmp()} the matrix element for the specific
configuration is evaluated. In the next lines the
individual contributions to the amplitude are retrieved from the
{\tt NGluon} class and compared with the results as shown in 
\Ref{Giele:2008bc}.
Compiling the file {\tt NGluon-demo.cpp} this test can be run through the
command line option \verb#--GZcheck#. The result from the comparison will
be printed on the screen. An example run is shown in the appendix \ref{sec:GZdemo}.
For small number of gluons we find good agreement with \Ref{Giele:2008bc}.
For the examples with a larger number
of gluons the agreement is getting worse. The examples for high 
multiplicities were calculated in extended precision in \Ref{Giele:2008bc}. We
also tried to switch to extended precision, however we do not observe
a significant improvement. We believe that this is due to the fact that
the momentum configuration is only given with double accuracy. Since
momentum conservation and on-shellness are satisfied only to 15 digits 
switching to extended accuracy does not give significant improvement in this case.
Using the option \verb#--GKMcheck# the program will compare with
results published in \Ref{Giele:2008ve}. The sample output is shown in 
appendix \ref{sec:GKMdemo}. In case of the 5 gluon amplitude we
observed a discrepancy for the helicity configurations $--+++$ and
$-+-++$. We believe that this is due to a mismatch in the helicity
labeling since we get agreement when we flip the helicities to 
$++---$ and $+-+--$.

\subsection{Accuracy}
\subsubsection{The scale test}
\begin{figure}[htbp]
  \begin{center}
    \leavevmode
    \includegraphics[width=\textwidth]{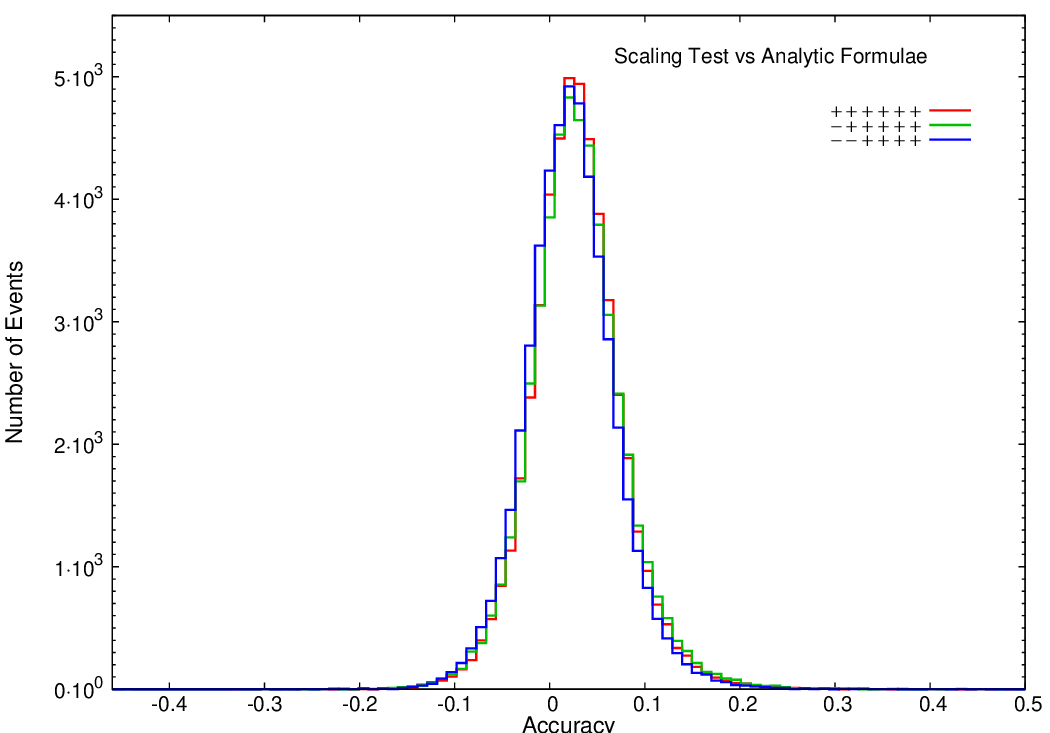}
    \caption{Test of the accuracy estimated from the scale test.}
    \label{fig:scaletest}
  \end{center}
\end{figure}
Although {\tt NGluon} can be compiled to work with extended
precision all checks in this section are obtained in double precision.
This gives a direct measure for the numerical stability of the
program. In the practical application one may resort to extended
floating point precision, to recalculate phase space points which
could not be calculated in double precision.
Since some of the checks make use of the scale test described in the
previous section we first assess the question how reliable this test
is. To do so we study a 6-gluon amplitude, where analytic results for
different helicity configurations are available \cite{Mahlon:1993si,Bern:2005hs,
Bern:2005cq,Forde:2005hh,Badger:2008cm}. 
For the helicity configurations $(+)^6$, $-(+)^5$ and $--(+)^4$ 50000 phase
space points passing the same cuts as used in \Ref{Giele:2008ve} were
generated. The kinematic cut is applied to	 restrict the phase space
points to the ``physical region'' avoiding soft and collinear
configurations which may introduce further numerical
instabilities. Note that in a real application the IR safe jet
algorithm would provide this cut. For the 50000 phase space points the 
matrix elements were calculated using {\tt NGluon } with the scale
test switched on. 
In addition also the analytic formulae are used to calculate the
matrix element. We estimate the
absolute uncertainty  $\delta_s$ from the scale test by taking the
difference of the two results (after rescaling). 
The absolute uncertainty estimated from
the comparison with the analytic result is defined as the difference
of the numerical result and the result obtained from the evaluation of
the analytic formula. The absolute uncertainties are converted to
relative ones and the absolute value is taken. The logarithm of the
relative uncertainty provides an easy measure for the accuracy:
\begin{equation}
  d_s = \log\left( 2 \frac{\delta_s}{ A_1+A_2}\right)
  = \log\left( 2\frac{A_1 - A_2 }{ A_1+A_2}\right),
\end{equation}
and
\begin{equation}
  d_a = \log\left( 2 \frac{\delta_a }{ A_1+A_a}\right)
  = \log\left( 2 \frac{A_1 - A_a }{ A_1+A_a}\right)
\end{equation}
where $A_i$ are the two numerical results and $A_a$ the one from the
evaluation of the analytic formula.
In the ideal case the two
uncertainties would be 100\% correlated. In \Fig{fig:scaletest}   
we show the distribution of
\begin{equation}
    \frac{d_s}{ d_a} - 1.
\end{equation}
As one can see most of the events are located close to zero. The
scale test gives thus a reliable estimate for the uncertainty. It is clear
that the two methods to assess the uncertainty will not return
precisely the same result. However, from the small width of the
distribution we conclude that the scale test can replace the analytic
comparison when no analytic results are available.
Inspecting \Fig{fig:scaletest} in detail we observe that the
distribution is slightly shifted to the right. This
shift could be due to the details of the floating point arithmetic
combined with the fact that we assume that the results based on the
analytic formulae are always correct. For events where we estimate an
accuracy of ten or even more digits the evaluation of the analytic
formula may also not be precise enough to compare with. 
Another effect might be that
accidentally we may estimate for a specific event a higher accuracy than
we actually have.  It is possible that just by chance a digit which
is already out of numerical control agrees. This could happen in the
comparison with the analytic result as well as in the scale
test. Due to the prejudice that the analytic results are always
correct this could lead to a shift in one direction.  

\subsubsection{$n$-point MHV amplitude}
\begin{figure}[htbp]
  \begin{center}
    \leavevmode
    \includegraphics[width=0.48\textwidth]{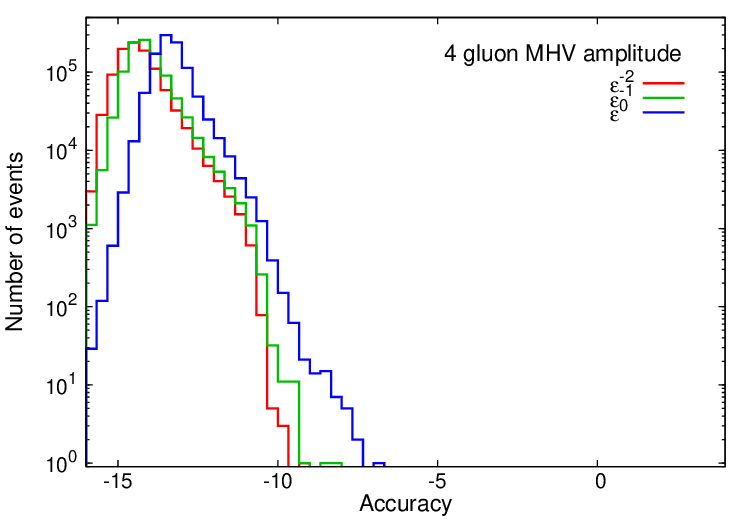}
    \includegraphics[width=0.48\textwidth]{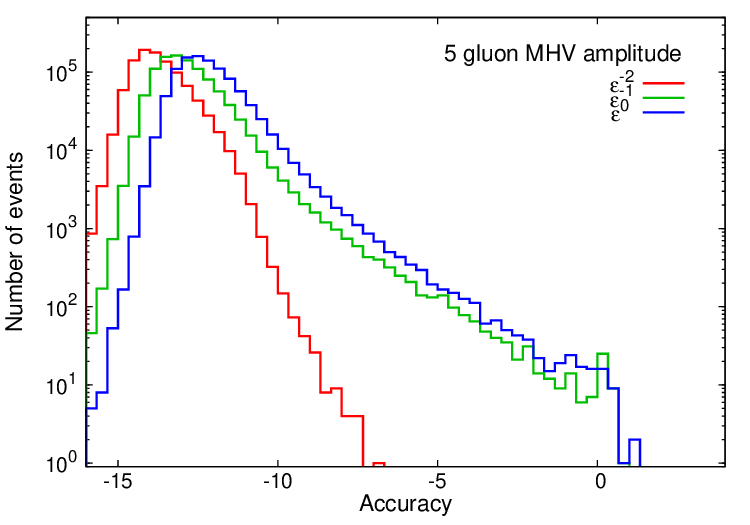}

     \includegraphics[width=0.48\textwidth]{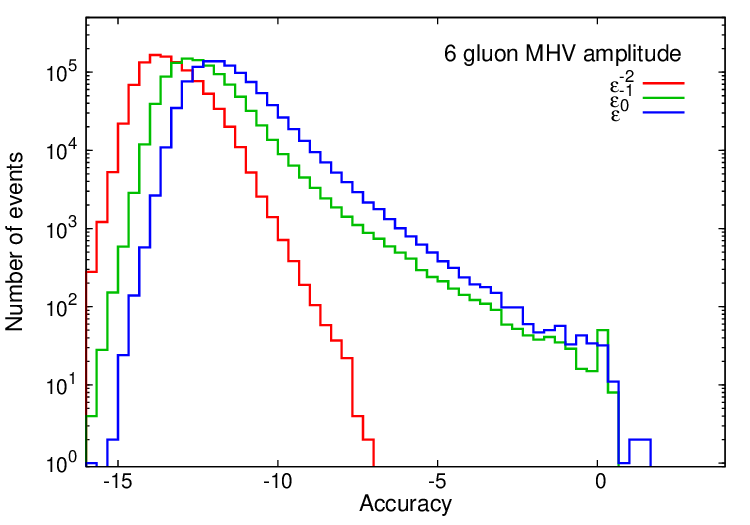}
     \includegraphics[width=0.48\textwidth]{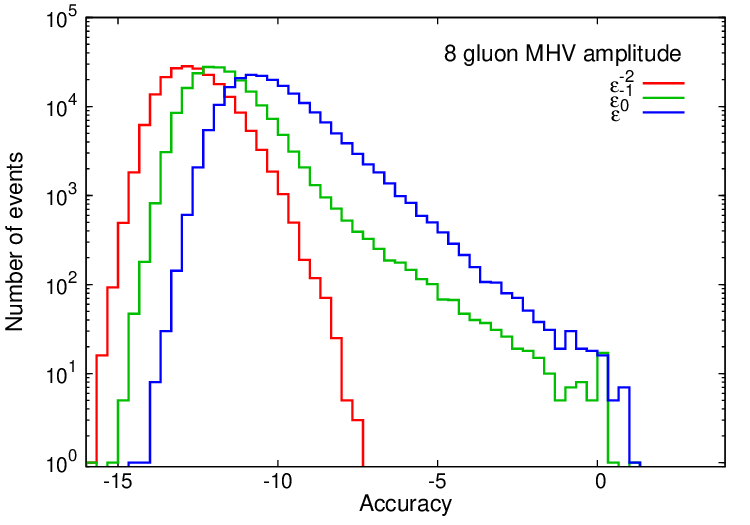}

     \includegraphics[width=0.48\textwidth]{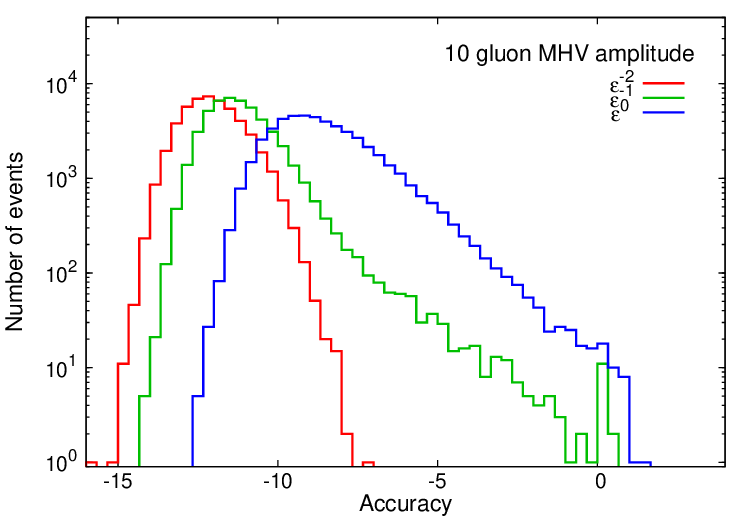}
     \includegraphics[width=0.48\textwidth]{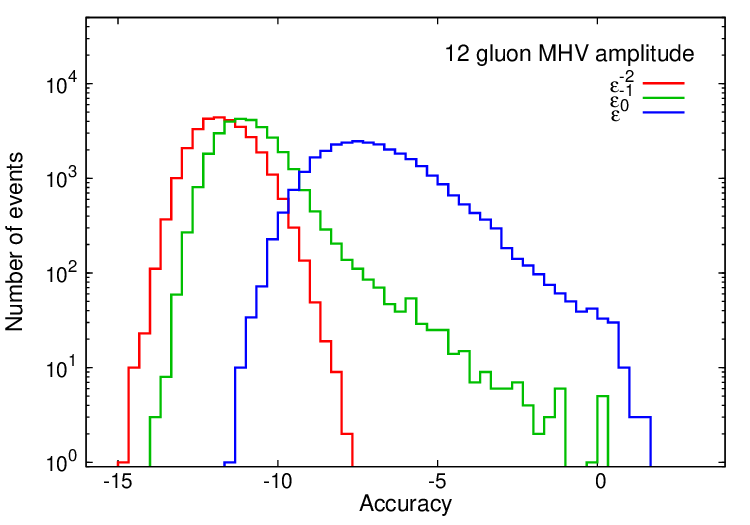}
    \caption{Accuracy for the $\frac{1}{ \epsilon^2}$-pole,
     $\frac{1}{\epsilon}$-pole and the finite part.}
    \label{fig:AccPlot}
  \end{center}
\end{figure}
\begin{figure}[htbp]
  \begin{center}
    \leavevmode
    \includegraphics[width=0.9\textwidth]{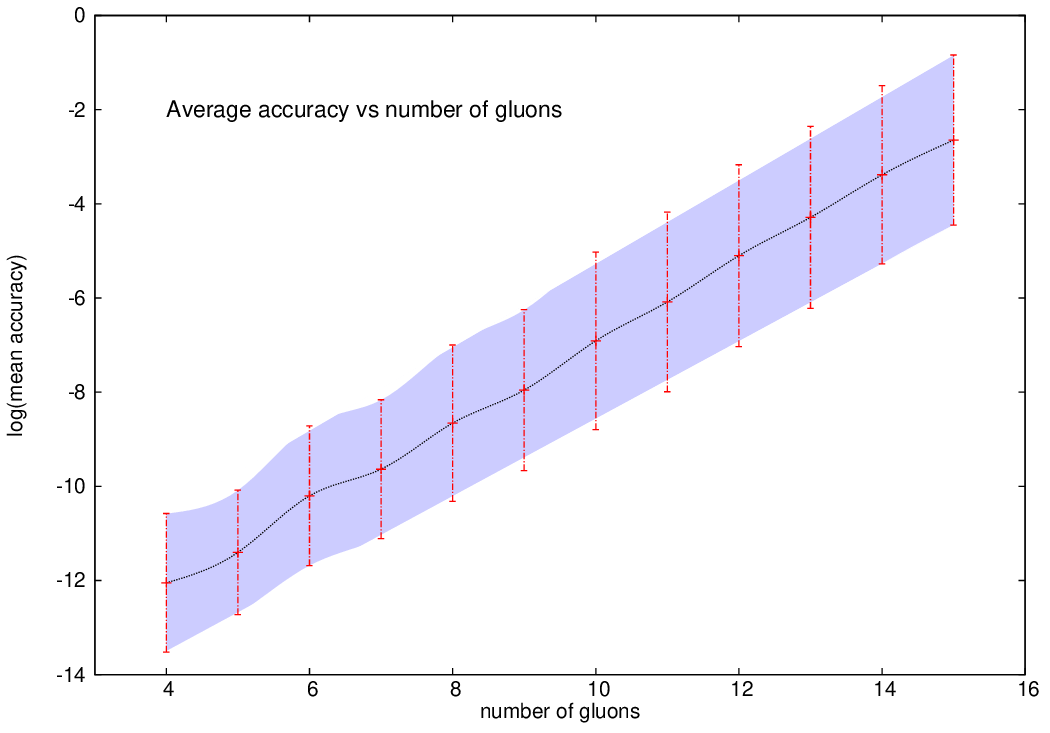}
    \caption{The average accuracy as estimated from the scale test as
    function of the number of gluons. The band gives a measure for the
    width  of the distribution}
    \label{fig:AverageAcc}
  \end{center}
\end{figure}
\begin{figure}[htbp]
  \begin{center}
    \leavevmode
    \includegraphics{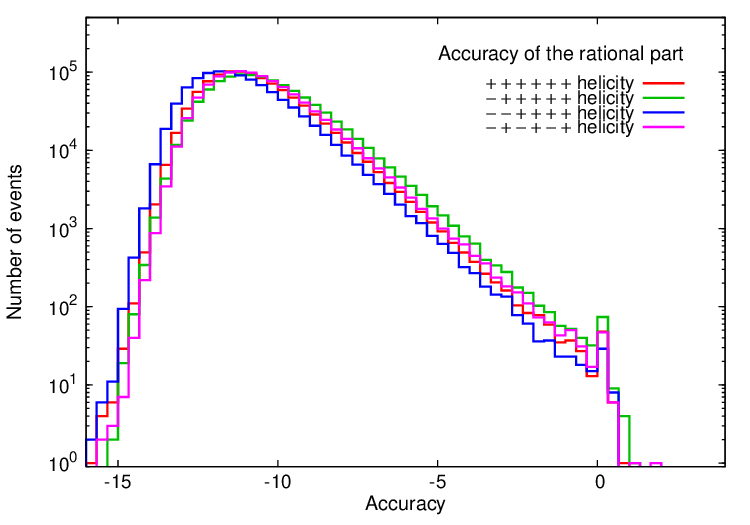}
    \caption{Accuracy of the rational part for the 6 gluon amplitude
      for different helicity configurations.}
    \label{fig:AccPlot6GKM}
  \end{center}
\end{figure}

To perform further checks on the correctness of the code and 
also the performance with
respect to numerical accuracy/stability we analysed the
accuracy for different phase space points for amplitudes of different 
complexity. More specifically, we consider the case of
$n=4,5,6,8,10,12$ gluons.
The number of phase space points for each case is fixed through the 
requirement that each test should be done
in less than a couple of hours. (For 4--6 gluons we used 1000000 
phase space
points, for 8 gluons we used 200000, for 10 gluons 50000 and for
12 gluons 30000 phase space points were used.)
We analysed the accuracy for the $1/\epsilon^2,
1/\epsilon$ poles as well as the accuracy for the finite part. The events were
binned according to the accuracy. We take again the logarithm of the
relative uncertainty as a measure for the accuracy. The result is shown
in \Fig{fig:AccPlot}. We considered the MHV amplitudes since analytic
results for the rational part 	exist. The rational part is numerically the most
complicated contribution. The accuracy can thus be taken as a
pessimistic point of view. 
In all cases --- even for high multiplicities --- we observe that the
leading IR singularities can be determined with high accuracy. The
accuracy is never worse than $-7$. For the 4 gluon amplitude we find
that the different contributions --- pole parts and finite parts ---
show similar behaviour. The accuracy is sufficiently good for most of
the phenomenological applications at the LHC (if not for all). However,
since in that case also analytic results are available this is not of
any practical use. Beyond $n=4$ the distributions show a
similar behaviour for different number of gluons. The $\frac{1}{\epsilon}$-poles
follow to some extend the finite parts. With increasing number of
gluons the peak of the distributions is shifted to the right. With
more gluons the computation is getting more involved, and the average
accuracy decreases. Due to the logarithmic scale the histograms may be 
misleading. We note that for the most complicated case shown in 
\Fig{fig:AccPlot} --- the 12 gluon amplitude --- only about 3\% of all
events have an accuracy above $-3$. In table
\Tab{tab:BadPointsFraction} 
we show the
fraction of events with an accuracy above $-3$.
\begin{table}[htbp]
  \begin{center}
    \leavevmode
    \begin{tabular}[h]{c|l}
      $n$ gluons & bad points [\%]\\
      \hline
      4 & --- \\
      5 & 0.03 \\
      6 & 0.06 \\
      8 & 0.2 \\
      10 & 0.8 \\
      12 & 3.
    \end{tabular}
    \caption{Fraction of events with an accuracy above $-3$.}
    \label{tab:BadPointsFraction}
  \end{center}
\end{table}
In all cases the fraction shown is evaluated for the $--(+)^{n-2}$ 
configuration. Considering different configurations may change the
fractions. We also observed that asking for an accuracy of $-4$ will
actually double the fraction of bad points. This would mean that for
the 12 gluon case, for example, about 6 \% of the events need to be
reevaluated with higher accuracy. 
In \Fig{fig:AverageAcc} we show the average accuracy evaluated for 
a fixed number of phase space points. The accuracy is evaluated using 
the scale test.
As we can see the accuracy is a linearly raising function of the
number of gluons. Starting at $n=4$  with an average accuracy of about
12 digits, the accuracy reaches $-4$ for about 14 gluons. For 14
gluons we have thus on average only $3-4$ digits which are
significant. To estimate whether the program can still be run mostly
with built-in double precision in addition to  the average accuracy
the width of the distribution is important. The width is
illustrated as a blue band. We observe in \Fig{fig:AverageAcc} that
the width increases as a function of the number of
gluons. However, the effect is only moderate in size.
Assuming that for most LHC applications 4 significant digits
should be sufficient we can conclude from 
\Fig{fig:AverageAcc} that up to 12--14 gluons the program may be 
used with only a small fraction of points requiring a recalculation
using extended precision.
\begin{figure}[htbp]
  \begin{center}
    \leavevmode
    \includegraphics{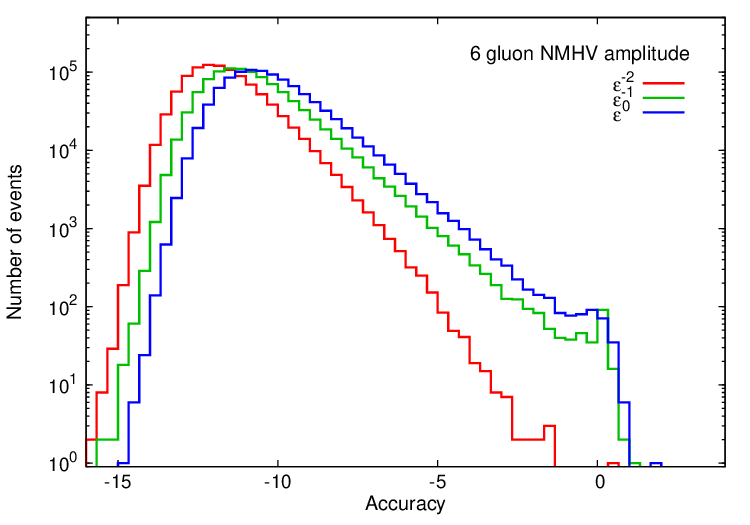}
    \caption{Accuracy for a NMHV amplitude for 6 gluons.}
    \label{fig:AccPlot6NMHV}
  \end{center}
\end{figure}
One may ask the question how much the findings on the accuracy shown
before depend on the specific helicity configuration. In 
\Fig{fig:AccPlot6NMHV} we show a similar plot as discussed before
but now for the NMHV configuration. We observe that leading
$1/\epsilon^2$ singularity is changed compared to what we have seen
before. Since the pole part is known analytically this has no
practical consequences.
Inspecting the accuracy of the $\frac{1}{\epsilon}$ pole as well as the
accuracy of the finite part we observe a behaviour similar to what
has been shown in \Fig{fig:AccPlot}. In \Fig{fig:AccPlot6GKM}
we show the accuracy of the rational part for different helicity 
configurations. Apart from minor differences at the right end of the 
histograms --- which may be due to statistical fluctuations --- we
observe that the behaviour is largely independent from the 
chosen helicity configuration. To good approximation we thus believe that our findings are to large
extend universal.

\subsection{Estimated speed}

\begin{figure}[htbp]
  \begin{center}
    \leavevmode
    \includegraphics[width=\textwidth]{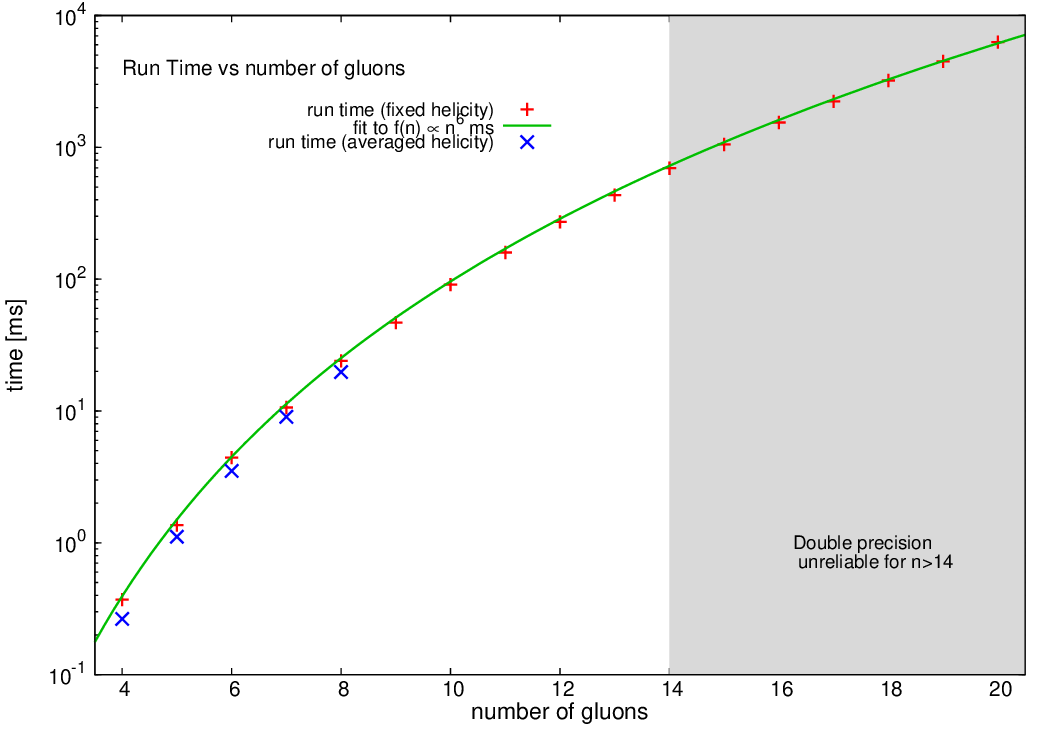}
    \caption{Runtime in milli seconds for the evaluation of one phase 
      space point. The red crosses show the runtime for the helicity
      configuration which we consider the worst case. For a comparison
      we also show for low multiplicities an average over all
      possible helicity configurations (blue crosses).}
    \label{fig:Runtime}
  \end{center}
\end{figure}
In \Fig{fig:Runtime} we show the runtime in milli seconds required to
evaluate one phase space point. The values were obtained by averaging
over several evaluations to avoid large fluctuations. 
As hardware we used an Intel Quad Core CPU (Q9650) with a
frequency of 3 GHz. As an operating system we used Scientific 
Linux Version 5.2. The program has been compiled with the GNU
compiler suite version 4.1.
The red crosses
show the runtime for the evaluation of $(-+)^{n/2}$
amplitudes. In principle, the runtime can depend on the helicity
configuration since for specific cases some of the Born amplitudes
entering the calculation may
vanish. The helicity configuration shown here should correspond to the worst
case. For low multiplicities we have also checked the runtime by
averaging over all possible helicity configurations. We find indeed
that the average runtime obtained in this way are smaller than the ones
measured for the $(-+)^{n/2}$ amplitudes. 
Since with increasing multiplicity the
number of different helicity configurations becomes quite large, we have
restricted this analysis to $n\le 8$. 

In Ref. \cite{Giele:2008bc} the authors showed that the computational costs behave as $n^9$.
Due to the cache system used for the evaluation of tree amplitudes we expect to find $n^8$. In
\Fig{fig:Runtime} the runtime is shown as function of the number of gluons. It turns out that for
$n=20$ the asymptotic regime of large $n$ is not yet reached and we observe, in contrast to the
expectation, a scaling like $n^6$. Extending the analysis to extreme values for $n$ we find indeed
that the scaling approaches $n^8$.  In particular we have checked that the pentagon contributions
follow the $n^8$ behaviour as expected.  However the numerical coefficient in front of
the $n^8$ polynomial is supressed so we do not observe the $n^8$ behaviour for small $n$ when
this contribution is combined with the remaing ones. Since running in double precision the results
for $n>20$ are unreliable we have restricted ourselves to 20 gluons in \Fig{fig:Runtime}. One should
thus not take the $n^6$ behaviour as asymptotic scaling for large $n$.  The $n^6$ behaviour is
represented in \Fig{fig:Runtime} by the solid green line.  Note that the region $n>14$ is shown for
illustrative purpose only. In this region the numerical accuracy is so bad that a significant
fraction of phase space points would require a re-evaluation using extended floating point
arithmetic.

The data used in \Fig{fig:Runtime} can be obtained by running the
program {\tt NGluon-demo} with the option {\verb#--runtime#}. We believe
that further improvements are possible. In particular {\tt NGluon} is
based on the evaluation of tree amplitudes using the Berends-Giele
recursion \cite{Berends:1987me}. In the implementation of the tree
amplitudes we have not used the fact that we are working in the
helicity basis. This could give a further improvement. We stress that
in our code no use of explicit analytic formulae for specific
helicity configurations is made. This could give a further improvement. 

\section{Conclusions} 
\label{sec:conclusions}
We have presented a purely numerical implementation of one-loop
$n$-gluon amplitudes. The implementation makes use of the generalised unitarity
method developed recently by various groups. We have performed
extensive checks on the numerical accuracy as well as the performance
with respect to the computing time. The program has passed all these checks,
is easy to use and shows a good behaviour with respect to the evaluation speed.
We believe that the program can be used to obtain reliable predictions
for the LHC. To the best of our knowledge the program
is the only available public code which allows the numerical
evaluation of arbitrary $n$-gluon one-loop amplitudes without requiring additional external
input. Apart from the physical application, the package
may serve as a reference implementation for future
developments. In particular, it would be extremely useful to extend to full QCD processes.
The extension to include fermionic loops is foreseen for
the next major release. 

\subsection*{Acknowledgments}
We would like thank Pierpaolo Mastrolia and Francesco Tramontano for illuminating discussions. We
are also grateful to Valery Yundin for advice and testing of the code.
This work is supported in part by the Deutsche Forschungsgemeinschaft
through the Transregional Collaborative Research Centre SFB-TR9 ``Computational
Particle Physics''and the Research Training Group (GK1504)
"Mass, Spectrum, Symmetry, Particle Physics in the Era of the Large Hadron Collider"
GK1504. In addition we acknowledge support   
from the Helmholtz Alliance {\it ``Physics at the Terascale''} contract VH-HA-101. The work of SB
has been supported in part by Danish Natural Science Reseach Council grant 10-084954.

\appendix
\section{Sample output}
\subsection{Comparison with Giele, Zanderighi}
\label{sec:GZdemo}
Running the sample application {\tt NGluon-demo} with the option
\verb#--GZcheck# should produce output similar to what is listed
below:{\small
\begin{verbatim}
# INTEGRALS: FF [1] and QCDLoop [2] are used to calculate the
# INTEGRALS: scalar one-loop integrals
# INTEGRALS: [1] van Oldenborgh: FF: A Package To Evaluate One Loop Feynman Diagrams
# INTEGRALS:     Comput.Phys.Commun.66:1-15,1991
# INTEGRALS: [2] R.Keith Ellis, Giulia Zanderighi,  Scalar one-loop integrals for QCD, 
# INTEGRALS:     JHEP 0802:002,2008


---------------------------------------------------------------
Numerical comparison with values published in: 
Giele,Zanderighi: 
On the Numerical Evaluation of One-Loop Amplitudes:
The Gluonic Case.
JHEP 0806:038,2008. 
---------------------------------------------------------------


---------------------------------------------------------------
#Number of gluons = 6
---------------------------------------------------------------
# ONELOOP: Renormalization scale set to mu = 3.600000000000000e+01
Helicities: ++++++
tree           : 2.449772238656663e-15
tree     (GZ)  : 0.000000000000000e+00
|Aeps2|        : 0.000000000000000e+00
|Aeps2|  (GZ)  : 0.000000000000000e+00
|Aeps1|        : 0.000000000000000e+00
|Aeps1|  (GZ)  : 0.000000000000000e+00
|Afinite|      : 5.298064836614244e-01
|Afinite|(GZ)  : 5.298064836438550e-01
Helicities: -+++++
tree           : 4.452705681580742e-14
tree     (GZ)  : 0.000000000000000e+00
|Aeps2|        : 7.016762080329839e-13
|Aeps2|  (GZ)  : 0.000000000000000e+00
|Aeps1|        : 7.119361750603557e-12
|Aeps1|  (GZ)  : 0.000000000000000e+00
|Afinite|      : 3.259967054240548e+00
|Afinite|(GZ)  : 3.259967054272360e+00
Helicities: --++++
tree           : 2.849128165044318e+01
tree     (GZ)  : 2.849128165044320e+01
|Aeps2|        : 1.709476899026886e+02
|Aeps2|  (GZ)  : 1.709476899026590e+02
|Aeps1|        : 6.145908783806966e+02
|Aeps1|  (GZ)  : 6.145908783763970e+02
|Afinite|      : 1.373747535025069e+03
|Afinite|(GZ)  : 1.373747535008280e+03
Helicities: -+-+-+
tree           : 3.138715395008066e+00
tree     (GZ)  : 3.138715395008080e+00
|Aeps2|        : 1.883229237005350e+01
|Aeps2|  (GZ)  : 1.883229237004850e+01
|Aeps1|        : 6.770582929013867e+01
|Aeps1|  (GZ)  : 6.770582928695769e+01
|Afinite|      : 1.510439503524497e+02
|Afinite|(GZ)  : 1.510439503379470e+02
Helicities: +-+-+-
tree           : 3.138715395008066e+00
tree     (GZ)  : 3.138715395008080e+00
|Aeps2|        : 1.883229237005516e+01
|Aeps2|  (GZ)  : 1.883229237004850e+01
|Aeps1|        : 6.770582928536295e+01
|Aeps1|  (GZ)  : 6.770582928695769e+01
|Afinite|      : 1.537801016025857e+02
|Afinite|(GZ)  : 1.537801014159860e+02
---------------------------------------------------------------
#Number of gluons = 7
---------------------------------------------------------------
# ONELOOP: Renormalization scale set to mu = 4.900000000000000e+01
Helicities: +++++++
tree           : 1.164850097354693e-15
tree     (GZ)  : 0.000000000000000e+00
|Aeps2|        : 0.000000000000000e+00
|Aeps2|  (GZ)  : 0.000000000000000e+00
|Aeps1|        : 0.000000000000000e+00
|Aeps1|  (GZ)  : 0.000000000000000e+00
|Afinite|      : 3.101695333690422e-01
|Afinite|(GZ)  : 3.101695334831830e-01
Helicities: -++++++
tree           : 1.414070990000164e-15
tree     (GZ)  : 0.000000000000000e+00
|Aeps2|        : 2.393016434615393e-14
|Aeps2|  (GZ)  : 0.000000000000000e+00
|Aeps1|        : 1.074107565329303e-13
|Aeps1|  (GZ)  : 0.000000000000000e+00
|Afinite|      : 1.920528150920153e-01
|Afinite|(GZ)  : 1.920528147653950e-01
Helicities: --+++++
tree           : 2.106612834594481e+00
tree     (GZ)  : 2.106612834594490e+00
|Aeps2|        : 1.474628984216144e+01
|Aeps2|  (GZ)  : 1.474628984216140e+01
|Aeps1|        : 4.850089396312033e+01
|Aeps1|  (GZ)  : 4.850089396312130e+01
|Afinite|      : 8.731521551316330e+01
|Afinite|(GZ)  : 8.731521551386510e+01
Helicities: -+-+-+-
tree           : 1.101865680944418e-01
tree     (GZ)  : 1.101865680944420e-01
|Aeps2|        : 7.713059766611572e-01
|Aeps2|  (GZ)  : 7.713059766610950e-01
|Aeps1|        : 2.536843489960513e+00
|Aeps1|  (GZ)  : 2.536843489960750e+00
|Afinite|      : 5.933610502627595e+00
|Afinite|(GZ)  : 5.933610502945470e+00
Helicities: +-+-+-+
tree           : 1.101865680944418e-01
tree     (GZ)  : 1.101865680944420e-01
|Aeps2|        : 7.713059766609913e-01
|Aeps2|  (GZ)  : 7.713059766610950e-01
|Aeps1|        : 2.536843489960436e+00
|Aeps1|  (GZ)  : 2.536843489960750e+00
|Afinite|      : 6.042012410247624e+00
|Afinite|(GZ)  : 6.042012409916140e+00
...
\end{verbatim}

The lines marked with (GZ) correspond to the results as taken from 
\Ref{Giele:2008bc}.

\subsection{Comparison with Giele, Kunszt, Melnikov}
\label{sec:GKMdemo}
Running the sample program with option \verb#--GKMcheck# should
produce results similar to what is listed below:
{\small
\begin{verbatim}
# INTEGRALS: FF [1] and QCDLoop [2] are used to calculate the
# INTEGRALS: scalar one-loop integrals
# INTEGRALS: [1] van Oldenborgh: FF: A Package To Evaluate One Loop Feynman Diagrams
# INTEGRALS:     Comput.Phys.Commun.66:1-15,1991
# INTEGRALS: [2] R.Keith Ellis, Giulia Zanderighi,  Scalar one-loop integrals for QCD, 
# INTEGRALS:     JHEP 0802:002,2008


---------------------------------------------------------------
 Cross checking results of GKM [arXiv:0801.2237] 
---------------------------------------------------------------
---------------------------------------------------------------
4-point helicity amplitudes 
---------------------------------------------------------------
Helicities: ++++
amp      : 3.33333333e-01
amp(GKM) : 3.33330000e-01
Helicities: -+++
amp      : 7.50000000e-01
amp(GKM) : 7.50000000e-01
Helicities: --++
amp      : 2.75849329e+00
amp(GKM) : 2.75849000e+00
IR-pole      : 4.89407794e+00
IR-pole(ana) : 4.89407794e+00
Helicities: -+-+
amp      : 4.17948712e+00
amp(GKM) : 4.17948834e+00
IR-pole      : 4.89407794e+00
IR-pole(ana) : 4.89407794e+00
---------------------------------------------------------------
5-point helicity amplitudes 
---------------------------------------------------------------
Helicities: +++++
amp      : 6.61487185e-01
amp(GKM) : 6.61484296e-01
Helicities: -++++
amp      : 8.40420360e-01
amp(GKM) : 8.40440985e-01
Helicities: --+++
amp      : 9.41625089e+00
amp(GKM) : 8.39210351e+00
IR-pole      : 7.35468951e+00
IR-pole(ana) : 7.35468951e+00
amp(++---) 8.39210347e+00
Helicities: -+-++
amp      : 7.06950047e+00
amp(GKM) : 8.06284942e+00
IR-pole      : 7.35468951e+00
IR-pole(ana) : 7.35468951e+00
amp(+-+--) 8.06284951e+00
---------------------------------------------------------------
6-point helicity amplitudes 
---------------------------------------------------------------
Helicities: ++++++
amp      : 5.29806483e-01
amp(GKM) : 5.29806465e-01
Helicities: -+++++
amp      : 3.25996705e+00
amp(GKM) : 3.25996706e+00
Helicities: --++++
amp      : 9.73370506e+00
amp(GKM) : 9.73370449e+00
IR-pole      : 2.03178718e+00
IR-pole(ana) : 2.03178718e+00
Helicities: -+-+++
amp      : 9.28315860e+00
amp(GKM) : 9.28315867e+00
IR-pole      : 2.03178718e+00
IR-pole(ana) : 2.03178718e+00
Helicities: -++-++
amp      : 1.34373214e+01
amp(GKM) : 1.34373207e+01
IR-pole      : 2.03178718e+00
IR-pole(ana) : 2.03178718e+00
Helicities: ---+++
amp      : 1.78047530e+01
amp(GKM) : 1.78047526e+01
IR-pole      : 2.03178718e+00
IR-pole(ana) : 2.03178718e+00
Helicities: --+-++
amp      : 1.23425461e+01
amp(GKM) : 1.23425455e+01
IR-pole      : 2.03178718e+00
IR-pole(ana) : 2.03178718e+00
Helicities: -+-+-+
amp      : 1.48181613e+01
amp(GKM) : 1.48181614e+01
IR-pole      : 2.03178718e+00
IR-pole(ana) : 2.03178718e+00


---------------------------------------------------------------
  all checks with IR/UV poles and JHEP 0806:038,2008 passed. 
---------------------------------------------------------------


# Time used for this run: 4.00020000e-02
\end{verbatim}
}
We note that for the helicity configurations $--+++$ and $-+-++$
we disagree with \Ref{Giele:2008ve}. However, we observed that
switching to $++---$ and $+-+--$ we find agreement.
%
%

\cleardoublepage
%
%

{\footnotesize


\begin{thebibliography}{10}

\bibitem{Bredenstein:2009aj}
A.~Bredenstein, A.~Denner, S.~Dittmaier, and S.~Pozzorini,
\newblock Phys.Rev.Lett. {\bf 103}, 012002 (2009), arXiv:arXiv:0905.0110.

\bibitem{Dittmaier:2007th}
S.~Dittmaier, S.~Kallweit, and P.~Uwer,
\newblock Phys.Rev.Lett. {\bf 100}, 062003 (2008), arXiv:arXiv:0710.1577.

\bibitem{Dittmaier:2007wz}
S.~Dittmaier, P.~Uwer, and S.~Weinzierl,
\newblock Phys.Rev.Lett. {\bf 98}, 262002 (2007), arXiv:hep-ph/0703120.

\bibitem{Binoth:2010ra}
SM and NLO Multileg Working Group, J.~Andersen {\em et~al.},
\newblock (2010), arXiv:1003.1241.

\bibitem{Bern:1994zx}
Z.~Bern, L.~J. Dixon, D.~C. Dunbar, and D.~A. Kosower,
\newblock Nucl. Phys. {\bf B425}, 217 (1994), arXiv:hep-ph/9403226.

\bibitem{Bern:1994cg}
Z.~Bern, L.~J. Dixon, D.~C. Dunbar, and D.~A. Kosower,
\newblock Nucl. Phys. {\bf B435}, 59 (1995), arXiv:hep-ph/9409265.

\bibitem{Britto:2004nc}
R.~Britto, F.~Cachazo, and B.~Feng,
\newblock Nucl. Phys. {\bf B725}, 275 (2005), arXiv:hep-th/0412103.

\bibitem{Ossola:2006us}
G.~Ossola, C.~G. Papadopoulos, and R.~Pittau,
\newblock Nucl. Phys. {\bf B763}, 147 (2007), arXiv:hep-ph/0609007.

\bibitem{Forde:2007mi}
D.~Forde,
\newblock Phys. Rev. {\bf D75}, 125019 (2007), arXiv:0704.1835.

\bibitem{Bern:2005hs}
Z.~Bern, L.~J. Dixon, and D.~A. Kosower,
\newblock Phys.Rev. {\bf D71}, 105013 (2005), arXiv:hep-th/0501240.

\bibitem{Bern:2005ji}
Z.~Bern, L.~J. Dixon, and D.~A. Kosower,
\newblock Phys.Rev. {\bf D72}, 125003 (2005), arXiv:hep-ph/0505055.

\bibitem{Bern:2005cq}
Z.~Bern, L.~J. Dixon, and D.~A. Kosower,
\newblock Phys.Rev. {\bf D73}, 065013 (2006), arXiv:hep-ph/0507005.

\bibitem{Berger:2006ci}
C.~F. Berger, Z.~Bern, L.~J. Dixon, D.~Forde, and D.~A. Kosower,
\newblock Phys.Rev. {\bf D74}, 036009 (2006), arXiv:hep-ph/0604195.

\bibitem{Berger:2006vq}
C.~F. Berger, Z.~Bern, L.~J. Dixon, D.~Forde, and D.~A. Kosower,
\newblock Phys.Rev. {\bf D75}, 016006 (2007), arXiv:hep-ph/0607014.

\bibitem{Anastasiou:2006jv}
C.~Anastasiou, R.~Britto, B.~Feng, Z.~Kunszt, and P.~Mastrolia,
\newblock Phys. Lett. {\bf B645}, 213 (2007), arXiv:hep-ph/0609191.

\bibitem{Anastasiou:2006gt}
C.~Anastasiou, R.~Britto, B.~Feng, Z.~Kunszt, and P.~Mastrolia,
\newblock JHEP {\bf 03}, 111 (2007), arXiv:hep-ph/0612277.

\bibitem{Giele:2008ve}
W.~T. Giele, Z.~Kunszt, and K.~Melnikov,
\newblock JHEP {\bf 04}, 049 (2008), arXiv:0801.2237.

\bibitem{Ossola:2008xq}
G.~Ossola, C.~G. Papadopoulos, and R.~Pittau,
\newblock JHEP {\bf 05}, 004 (2008), arXiv:0802.1876.

\bibitem{Badger:2008cm}
S.~D. Badger,
\newblock JHEP {\bf 01}, 049 (2009), arXiv:0806.4600.

\bibitem{Ellis:2007br}
R.~K. Ellis, W.~T. Giele, and Z.~Kunszt,
\newblock JHEP {\bf 03}, 003 (2008), arXiv:0708.2398.

\bibitem{Berger:2008sj}
C.~F. Berger {\em et~al.},
\newblock Phys. Rev. {\bf D78}, 036003 (2008), arXiv:0803.4180.

\bibitem{Giele:2008bc}
W.~Giele and G.~Zanderighi,
\newblock JHEP {\bf 0806}, 038 (2008), arXiv:arXiv:0805.2152.

\bibitem{Giele:2009ui}
W.~Giele, Z.~Kunszt, and J.~Winter,
\newblock (2009), arXiv:0911.1962.

\bibitem{Lazopoulos:2008ex}
A.~Lazopoulos,
\newblock (2008), arXiv:arXiv:0812.2998.

\bibitem{Lazopoulos:2009zn}
A.~Lazopoulos,
\newblock (2009), arXiv:arXiv:0911.5241.

\bibitem{Berger:2009ep}
C.~F. Berger {\em et~al.},
\newblock Phys. Rev. {\bf D80}, 074036 (2009), arXiv:0907.1984.

\bibitem{Berger:2009zg}
C.~F. Berger {\em et~al.},
\newblock Phys. Rev. Lett. {\bf 102}, 222001 (2009), arXiv:0902.2760.

\bibitem{Berger:2010vm}
C.~F. Berger {\em et~al.},
\newblock (2010), arXiv:1004.1659.

\bibitem{Berger:2010zx}
C.~Berger {\em et~al.},
\newblock (2010), arXiv:arXiv:1009.2338,
\newblock * Temporary entry *.

\bibitem{Melia:2010bm}
T.~Melia, K.~Melnikov, R.~Rontsch, and G.~Zanderighi,
\newblock (2010), arXiv:arXiv:1007.5313.

\bibitem{Melnikov:2010iu}
K.~Melnikov and M.~Schulze,
\newblock Nucl.Phys. {\bf B840}, 129 (2010), arXiv:arXiv:1004.3284.

\bibitem{Melnikov:2009wh}
K.~Melnikov and G.~Zanderighi,
\newblock Phys.Rev. {\bf D81}, 074025 (2010), arXiv:arXiv:0910.3671.

\bibitem{Melnikov:2009dn}
K.~Melnikov and M.~Schulze,
\newblock JHEP {\bf 0908}, 049 (2009), arXiv:arXiv:0907.3090.

\bibitem{Ellis:2009zw}
R.~Ellis, K.~Melnikov, and G.~Zanderighi,
\newblock JHEP {\bf 0904}, 077 (2009), arXiv:arXiv:0901.4101.

\bibitem{Bevilacqua:2010ve}
G.~Bevilacqua, M.~Czakon, C.~Papadopoulos, and M.~Worek,
\newblock Phys.Rev.Lett. {\bf 104}, 162002 (2010), arXiv:arXiv:1002.4009.

\bibitem{Bevilacqua:2009zn}
G.~Bevilacqua, M.~Czakon, C.~Papadopoulos, R.~Pittau, and M.~Worek,
\newblock JHEP {\bf 0909}, 109 (2009), arXiv:arXiv:0907.4723.

\bibitem{Ossola:2007ax}
G.~Ossola, C.~G. Papadopoulos, and R.~Pittau,
\newblock JHEP {\bf 03}, 042 (2008), arXiv:0711.3596.

\bibitem{Mastrolia:2010nb}
P.~Mastrolia, G.~Ossola, T.~Reiter, and F.~Tramontano,
\newblock JHEP {\bf 1008}, 080 (2010), arXiv:arXiv:1006.0710.

\bibitem{'tHooft:1978xw}
G.~'t~Hooft and M.~Veltman,
\newblock Nucl.Phys. {\bf B153}, 365 (1979).

\bibitem{Denner:1991qq}
A.~Denner, U.~Nierste, and R.~Scharf,
\newblock Nucl.Phys. {\bf B367}, 637 (1991),
\newblock Dedicated to M. Veltman on occasion of his 60th birthday.

\bibitem{Bern:1993kr}
Z.~Bern, L.~J. Dixon, and D.~A. Kosower,
\newblock Nucl.Phys. {\bf B412}, 751 (1994), arXiv:hep-ph/9306240.

\bibitem{Ellis:2007qk}
R.~K. Ellis and G.~Zanderighi,
\newblock JHEP {\bf 02}, 002 (2008), arXiv:0712.1851.

\bibitem{vanOldenborgh:1989wn}
G.~van Oldenborgh and J.~Vermaseren,
\newblock Z.Phys. {\bf C46}, 425 (1990).

\bibitem{Ellis:2008qc}
R.~K. Ellis, W.~T. Giele, Z.~Kunszt, K.~Melnikov, and G.~Zanderighi,
\newblock JHEP {\bf 01}, 012 (2009), arXiv:0810.2762.

\bibitem{Hahn:1998yk}
T.~Hahn and M.~Perez-Victoria,
\newblock Comput.Phys.Commun. {\bf 118}, 153 (1999), arXiv:hep-ph/9807565.

\bibitem{vanHameren:2010cp}
A.~van Hameren,
\newblock (2010), arXiv:arXiv:1007.4716.

\bibitem{Berger:2009zb}
C.~F. Berger and D.~Forde,
\newblock Ann.Rev.Nucl.Part.Sci.  (2009), arXiv:arXiv:0912.3534.

\bibitem{Giele:1991vf}
W.~Giele and E.~Glover,
\newblock Phys.Rev. {\bf D46}, 1980 (1992).

\bibitem{Berends:1987me}
F.~A. Berends and W.~Giele,
\newblock Nucl.Phys. {\bf B306}, 759 (1988).

\bibitem{vanNeerven:1983vr}
W.~L. van Neerven and J.~A.~M. Vermaseren,
\newblock Phys. Lett. {\bf B137}, 241 (1984).

\bibitem{Ossola:2007bb}
G.~Ossola, C.~G. Papadopoulos, and R.~Pittau,
\newblock JHEP {\bf 07}, 085 (2007), arXiv:0704.1271.

\bibitem{delAguila:2004nf}
F.~del Aguila and R.~Pittau,
\newblock JHEP {\bf 07}, 017 (2004), arXiv:hep-ph/0404120.

\bibitem{Bern:1996ja}
Z.~Bern, L.~J. Dixon, D.~C. Dunbar, and D.~A. Kosower,
\newblock Phys. Lett. {\bf B394}, 105 (1997), arXiv:hep-th/9611127.

\bibitem{QD}
Y.~Hida, X.~S. Li, and D.~H. Bailey,
\newblock {Library for Double-Double and Quad-Double Arithmetic},
\newblock http://crd.lbl.gov/\u02dcdhbailey/mpdist, report LBNL-46996, 2008.

\bibitem{Mahlon:1993si}
G.~Mahlon,
\newblock Phys.Rev. {\bf D49}, 4438 (1994), arXiv:hep-ph/9312276.

\bibitem{Forde:2005hh}
D.~Forde and D.~A. Kosower,
\newblock Phys.Rev. {\bf D73}, 061701 (2006), arXiv:hep-ph/0509358.

\end{thebibliography}

}

\end{document}